\documentclass[11pt]{article}

\usepackage[pdftex]{graphicx}
\usepackage{amsmath}
\usepackage{epsf}
\usepackage[parfill]{parskip}
\usepackage[matrix,arrow]{xy}
\usepackage[usenames]{color}

\usepackage{epsfig}

\usepackage{amssymb,amsthm}
\usepackage{amscd}
\usepackage{amsfonts}
\usepackage{latexsym}
\usepackage{graphics}
\usepackage{color}
\usepackage{bbm}
\usepackage{bm}

\usepackage{xypic}
\usepackage{mathrsfs}

\input{xy}
\xyoption{all}

\input{epsf}

\usepackage[bookmarks=true]{hyperref}
\hypersetup{breaklinks=true} 

\let\fn\footnote
\renewcommand{\footnote}[1]{\linespread{1.1}\fn{#1}\linespread{1.29}}

\usepackage{fancyhdr}
\usepackage[left]{lineno}

\makeatletter\renewcommand{\section}{\@startsection
{section}{1}{\z@}{-3.5ex plus -1ex minus
    -.2ex}{2.3ex plus .2ex}{\bf }}
\makeatletter\renewcommand{\subsection}{\@startsection{subsection}{2}{\z@}{-3.25ex
plus -1ex minus
   -.2ex}{1.5ex plus .2ex}{\it }}
\makeatletter\renewcommand{\subsubsection}{\@startsection{subsubsection}{3}{-2.45ex}{-3.25ex
plus -1ex minus -.2ex}{1.5ex plus .2ex}{\it }}
\renewcommand{\thesection}{\arabic{section}.}
\renewcommand{\thesubsection}{\arabic{section}.\arabic{subsection}.}

\makeatletter\renewcommand{\subsubsection}
{\@startsection{subsubsection}{3}{\z@}{-3.25ex plus -1ex minus -.2ex}
{1.5ex plus .2ex}{\noindent\underline}}

\renewcommand{\theequation}{\thesection\arabic{equation}}
\makeatletter \@addtoreset{equation}{section}

\renewenvironment{thebibliography}[1]
     {\baselineskip=16pt plus 2pt minus 1pt
      \section*{\large\refname
        \@mkboth{\MakeUppercase\refname}{\MakeUppercase\refname}}%
     \list{\@biblabel{\@arabic\c@enumiv}}%
           {\settowidth\labelwidth{\@biblabel{#1}}%
            \leftmargin\labelwidth
            \advance\leftmargin\labelsep
            \@openbib@code
            \usecounter{enumiv}%
            \let\p@enumiv\@empty
            \renewcommand\theenumiv{\@arabic\c@enumiv}}%
      \sloppy
      \clubpenalty4000
      \@clubpenalty \clubpenalty
      \widowpenalty4000%
      \sfcode`\.\@m}

\newcommand{\acknowledgements}{\section*{Acknowledgements}
\addcontentsline{toc}{section}{\hspace{0.6cm}{\bf Acknowledgements}}}

\newcommand{\appendices}{\section*{Appendices}\setcounter{subsection}{0}\setcounter{equation}{0}\renewcommand{\thesubsection}{\Alph{subsection}.}
\renewcommand{\theequation}{\thesubsection\arabic{equation}}
\makeatletter \@addtoreset{equation}{subsection}
\addtocontents{toc}{\vspace{0.2cm}

{\bf Appendices}}
}

\usepackage{epsfig,rotating}
\usepackage{amsmath,amssymb}
\usepackage{amsfonts}
\usepackage[matrix,arrow]{xy}

\usepackage{mathrsfs}
\usepackage{bbm}
\usepackage{bm}
\usepackage[vcentermath,enableskew]{youngtab}

\def\slasha#1{\setbox0=\hbox{$#1$}#1\hskip-\wd0\hbox to\wd0{\hss\sl/\/\hss}}

\def\periodb#1{\setbox0=\hbox{$#1$}#1\hskip-\wd0\hbox to\wd0{-}}

\newcommand{\unit}{\mathbbm{1}}   			

\newcommand{\CA}{\mathcal{A}}    			

\newcommand{\CCA}{\mathscr{A}}

\newcommand{\CC}{\mathcal{C}}

\newcommand{\CCF}{\mathscr{F}}

\newcommand{\CH}{\mathcal{H}}

\newcommand{\CCI}{\mathscr{I}}

\newcommand{\CN}{\mathcal{N}}
\newcommand{\CCN}{\mathscr{N}}

\newcommand{\CQ}{\mathcal{Q}}

\newcommand{\CS}{\mathcal{S}}

\newcommand{\CV}{\mathcal{V}}

\newcommand{\CW}{\mathcal{W}}

\newcommand{\CZ}{\mathcal{Z}}
\newcommand{\CCZ}{\mathscr{Z}}

\newcommand{\frg}{\mathfrak{g}}				

\newcommand{\frh}{\mathfrak{h}}

\newcommand{\frM}{\mathfrak{M}}
\newcommand{\frS}{\mathfrak{S}}

\newcommand{\fru}{\mathfrak{u}}

\newcommand{\FR}{\mathbbm{R}}     			
\newcommand{\FC}{\mathbbm{C}}     			
\newcommand{\FT}{\mathbbm{T}}     			
\newcommand{\RZ}{\mathbbm{Z}}     			
\newcommand{\CPP}{{\mathbbm{C}P}}    			

\newcommand{\dd}{\mathrm{d}}     			
\newcommand{\de}{\mathrm{e}}     			
\newcommand{\di}{\mathrm{i}}     			
\newcommand{\eps}{{\varepsilon}}			

\newcommand{\Tr}{\,\mathrm{Tr}\,}

\newcommand{\asu}{\mathfrak{su}}
\newcommand{\aso}{\mathfrak{so}}

\newcommand{\sU}{\mathsf{U}}     			

\newcommand{\sSU}{\mathsf{SU}}

\newcommand{\sGL}{\mathsf{GL}}

\newcommand{\sSO}{\mathsf{SO}}

\newcommand{\sSpin}{\mathsf{Spin}}
\newcommand{\sEnd}{\mathsf{End}\,}
\newcommand{\sExt}{\mathsf{Ext}\,}
\newcommand{\sHom}{\mathsf{Hom}\,}

\newcommand{\remark}[1]{}     				
\def\tyng(#1){\hbox{\tiny$\yng(#1)$}}			
\def\tyoung(#1){\hbox{\tiny$\young(#1)$}}			

\newcommand{\mbf}[1]{{\boldsymbol {#1} }}

\newcommand{\ttA}{{\tt A}}

\newcommand{\beqa}{\begin{eqnarray}}
\newcommand{\eeqa}{\end{eqnarray}}
\newcommand{\beq}{\begin{equation}}
\newcommand{\eeq}{\end{equation}}

\begin{document}
\begin{titlepage}
\begin{flushright}
   EMPG--13--10 \\
\end{flushright}
\vskip 2.0cm
\begin{center}
{\LARGE \bf Reduced Chern-Simons Quiver Theories \\[3mm] and
  Cohomological 3-Algebra Models} \vskip 1.5cm {\Large Joshua
  DeBellis \ , \ Richard J. Szabo} \setcounter{footnote}{0}
\renewcommand{\thefootnote}{\arabic{thefootnote}} \vskip 1cm {\em Department of Mathematics\\
Heriot-Watt University\\
Colin Maclaurin Building, Riccarton, Edinburgh EH14 4AS, U.K.\\
Maxwell Institute for Mathematical Sciences, Edinburgh, U.K.\\
The Tait Institute, Edinburgh, U.K.}\\[5mm] {E-mail: {\ttfamily
  jd111@hw.ac.uk , R.J.Szabo@hw.ac.uk}} \vskip 1.1cm
\end{center}
\vskip 1.0cm
\begin{center}
{\bf Abstract}
\end{center}
\begin{quote}
We study the BPS spectrum and vacuum moduli spaces in dimensional
reductions of Chern-Simons-matter theories with $\CN\geqslant 2$
supersymmetry to zero dimensions. Our main
example is a matrix model version of the ABJM theory which we relate
explicitly to certain reduced 3-algebra models. We find the explicit
maps from Chern-Simons quiver matrix models to dual IKKT matrix
models. We address the problem of topologically twisting the ABJM
matrix model, and along the way construct a new twist of the IKKT
model. We construct a cohomological matrix model whose partition
function localizes onto a moduli space specified by 3-algebra
relations which live in the double of the conifold quiver. It computes an equivariant index enumerating framed BPS
states with specified R-charges which can be expressed as a combinatorial
sum over certain filtered pyramid partitions.
\end{quote}
\end{titlepage}

\tableofcontents

\allowdisplaybreaks


\bigskip

\section{Introduction and summary}

In this paper we study certain supersymmetric matrix models which are related to
Chern-Simons-matter theories in three dimensions with $\CN\geqslant2$
supersymmetry. Our motivation comes
from attempting to understand the spectrum of BPS states and the geometry of their
moduli spaces. A powerful tool for the enumeration of
supersymmetric vacua is provided by the Witten index since it is invariant under deformations of the continuous
parameters of the field theory. However, supersymmetric gauge theories
have much richer structures that are only partially captured by the
Witten index; to extract more information about the field theory, one
needs to exploit its symmetries. In three dimensions, a generalization of the Witten index is constructed using not only the
dilatation operator $H$, but also the $\sSO(2)$ angular momentum $J$ and the generators
$R_i$ of the
Cartan subalgebra of the R-symmetry group; schematically this refined index is given
by
\beq
\CCI(x,y,t) = \Tr_{\CH_{\rm BPS}}\, (-1)^F\, x^{H}\, y^{2J}\,
\prod_i\, t_i^{R_i}
\eeq
where the fugacities $x,y,t$ are inserted to resolve degeneracies. Like the Witten index, it
can be interpreted as a Feynman path integral with euclidean action
by compactifying the time direction on a circle $S^1$ with supersymmetric twisted
boundary conditions involving the $\sSO(2)$ rotation $J$ and the
global R-symmetry twists $R_i$; then $H$ is the generator of
translations along $S^1$ and $\CH_{\rm BPS}$ is the Hilbert
space of the theory with $\FR^2$ regarded as the spatial slice. In the
weak coupling limit $x\to0$ where the circle decompactifies, this theory reduces
to a supersymmetric quantum mechanics on the moduli space of BPS
solutions; in
this paper we aim to study an effective theory of these sorts of
reduced models.

As these field theories are expected to
flow to fixed points with superconformal symmetry, superconformal
field theories have found a distinctive place in the study of
supersymmetric gauge theories as well as in the AdS/CFT
correspondence. The pertinent index is then the superconformal index
which coincides with the partition function of the three-dimensional superconformal field
theory in radial quantization on $S^1\times S^2$. A large class of $\CN=2$
superconformal field theories in three dimensions arises as the
low-energy effective theory of 
multiple M2-branes probing a Calabi-Yau fourfold singularity. Using
AdS$_4$/CFT$_3$ duality, they
have a holographic dual description as M-theory on the product of
four-dimensional anti-de~Sitter space $AdS_4$ and a seven-dimensional
Sasaki-Einstein manifold $Y_7$ which is the base of the probed Calabi-Yau
fourfold cone. The simplest example is provided by the
theory of M2-branes at the orbifold singularity $\FC^4/\RZ_k$, which in the
low-energy limit is a Chern-Simons quiver gauge theory at level $k$ with $\CN=6$
supersymmetry~\cite{Aharony:2008ug,Distler:2008mk}; the holographic dual
theory is M-theory on $AdS_4\times (S^7/\RZ_k)$. For $k=1,2$, the
$\CN=6$ supersymmetry is enhanced to $\CN=8$, where the additional $\sSO(8)$ R-symmetry
generators are monopole
operators~\cite{Aharony:2008ug,Gustavsson:2009pm,Bashkirov:2010kz}. By
regarding $S^7$ as a circle bundle over $\CPP^3$ via the Hopf
fibration with
the orbifold action on the $S^1$ fibre, in
the Type~IIA frame the dual theory is supergravity on
$AdS_4\times\CPP^3$. This theory is the celebrated ABJM theory; in
this paper we mostly work with the more general ABJ
theory~\cite{Aharony:2008gk} with gauge group ${\sf U}(N_L)\times {\sf
  U}(N_R)$ which we still refer to as ``ABJM models'' for simplicity,
though we refer to the specialization $N_L=N_R=N$ as the ``ABJM
limit''. The superconformal index for ABJM theory is computed
in~\cite{Bhattacharya:2008bja,Kim:2009wb}, where it is matched with
that of the dual
Type~IIA theory and also M-theory in the large $N$ limit (see~\cite{Yokoyama:2011qu} for a review of superconformal indices in
three dimensions).

In this work we study in detail zero-dimensional reductions of
three-dimensional Chern-Simons quiver theories. This analysis is
partly inspired by the observation that the partition function of any $\CN\geqslant 2$ Chern-Simons-matter theory on
$S^3$ with no anomalous dimensions is given by a matrix integral over
the Cartan subalgebra of the gauge group~\cite{Kapustin:2009kz};
although our supersymmetric matrix models are structurally rather
different, we shall find that the matrix integrals reduce in the same
way. The result of~\cite{Kapustin:2009kz} can be used to confirm that the free energy of the infrared
$\CN=6$ superconformal field theory on $N$ M2-branes has the expected
$N^{3/2}$ scaling behaviour for large $N$, and it provides various
precise checks of the AdS$_4$/CFT$_3$ correspondence
(see~\cite{Marino:2011nm} for a review); more precisely, in the 't~Hooft limit $N,k\to\infty$ with $\lambda:=\frac Nk$ fixed, the
corresponding free energy grows like $(N\,
    k)^{3/2}$ times the volume of the $S^7/\RZ_k$ orbifold. More
    generally, for M-theory on $AdS_4\times Y_7$, where $Y_7$ is a seven-dimensional
Sasaki-Einstein space threaded by $N$ units of flux, the gravitational
free energy to leading order as $N\to\infty$ is
\beq
\CCF=N^{3/2}\, \sqrt{\frac{2\pi^6}{27\, {\rm Vol}(Y_7)}} \ ,
\eeq
where ${\rm Vol}(Y_7)$ is the riemannian volume of $Y_7$ with respect to
its Sasaki-Einstein metric; by the AdS/CFT correspondence, this result
can be
compared with computations in the dual Chern-Simons-matter field
theories which flow to superconformal fixed points. The partition functions of
three-dimensional $\CN=2$ superconformal field theories are similarly
computed in~\cite{Jafferis:2010un,Martelli:2011qj,Jafferis:2011zi};
their magnitudes are extremized at the superconformal R-charges of the
infrared conformal field theory. Here we take the point of view that supersymmetry allows one to relate the superconformal index to a
matrix integral, with action given by the reduction to zero dimensions
of the three-dimensional superconformal field theory.
Supersymmetric matrix models analogous to ours, obtained by dimensional reduction of
$\CN\geqslant 2$ Chern-Simons quiver gauge theories on $S^3$ to zero
dimensions, are considered in~\cite{Asano:2012gt,Honda:2012ni};
one-dimensional reductions of the ABJM theory are studied in~\cite{Berenstein:2008dc}.

Although much of our discussion will apply to reductions of generic Chern-Simons quiver
gauge theories, in the present paper we focus mostly on the partition functions of Chern-Simons-matter theories
with the highest amount $\CN\geqslant 6$ of supersymmetry, i.e. the
ABJM theory. In addition to the fact that they are important for the understanding of various aspects of
M-theory, it is these theories who
moduli spaces can be described as GIT quotients of bracket relations in certain
3-algebras. The Van Raamsdonk formulation of the ABJM model
\cite{VanRaamsdonk:2008ft} is a matrix field theory describing
stacks of M2-branes, and with a gauge group isomorphic to $ {\sf SU}(2)\times {\sf
  SU}(2)$ it is an $\mathcal{N}=8$ field theory which is equivalent to
the BLG theory~\cite{Bagger:2007jr,Gustavsson:2007vu}; the BLG model
is a three-dimensional $\mathcal{N}=8$ superconformal Chern-Simons
theory whose matter fields take values in a 3-Lie algebra with inner product compatible with the 3-bracket. A general $\mathcal{N}=6$ theory of multiple M2-branes was formulated
in \cite{Bagger:2008se} by relaxing the requirement of total
antisymmetry of the 3-bracket; then one may also reformulate
the ABJM theory with matter fields taking values in a particular hermitian
3-algebra. In this way one may study the novel geometry of the
supersymmetric gauge theory moduli spaces using 3-algebra
structures. This programme was initiated and studied
in~\cite{DeBellis:2010sy}, where dimensional reduction of the BLG
model was introduced and its BPS solutions were related to the
quantization of Nambu-Poisson manifolds, in the sense
of~\cite{DeBellis:2010pf}, which arise as worldvolume geometries of
M2-branes and M5-branes. In this paper we explore the relationship
between our Chern-Simons quiver matrix models and reduced 3-algebra
models, and exploit it in our studies of the vacuum moduli spaces.

Another reason for focusing on these particular classes of reductions of Chern-Simons quiver
theories is that these are the ones which bear an intimate
relationship to the better understood and behaved Yang-Mills matrix
models; in this paper we detail these matrix model relationships and
use them as guidance in our computations of partition functions. The ABJM
theory for $k=1$ is believed to have $\CN=8$ supersymmetry and to be
isomorphic to the infrared fixed point of maximally supersymmetric 
Yang-Mills theory in three dimensions. The partition functions for the
(mass
deformed) theories on $S^3$ are shown by~\cite{Kapustin:2010xq} to
agree, while their superconformal indices are matched
in~\cite{Gang:2011xp} and also with those of supergravity. The ABJM
model is also dual under mirror symmetry to $\CN=4$
$\sU(N)$ supersymmetric Yang-Mills theory coupled to a single
fundamental hypermultiplet and a single adjoint hypermultiplet.
Structural similarities between observables in $\CN=4$ supersymmetric
Yang-Mills theory and ABJM theory are also pointed out
in~\cite{Agarwal:2008pu,Wiegandt:2011uu}. In this paper we shall
exploit such equivalences as a duality between our reduced ABJM
model and the IKKT matrix model in ten dimensions~\cite{Ishibashi:1996xs}.
Reduced Yang-Mills theories play a central role in the nonperturbative
definitions of M-theory and superstring
theory~\cite{Banks:1996vh,Ishibashi:1996xs}. Yang-Mills matrix models
also provide a nonperturbative framework for emergent spacetime geometry and
noncommutative gauge
theories~\cite{Connes:1997cr,Aoki:1998vn,Aoki:1999vr,Ambjorn:1999ts}
(see~\cite{Szabo:2001kg,Steinacker:2010rh} for reviews); in
particular, the quantum geometries that arise as BPS solutions in the
reduced 3-Lie algebra model of~\cite{DeBellis:2010sy} are dual to
those of the IKKT model in ten dimensions. 
Investigations of the ABJM model as a new kind of matrix model for
M-theory, which is perhaps more fundamental than the BFSS matrix
theory~\cite{Banks:1996vh}, were carried out in~\cite{Mohammed:2010eb}.

The final simplifying aspect of this paper is that we wish to
formulate and solve topologically twisted models that properly
localize the dynamics onto the BPS moduli spaces; these are theories
that have a fermionic scalar symmetry which is a twisted version of
the supersymmetry of the original physical theory. While the IKKT
Yang-Mills matrix models can be deformed and solved using the powerful formalism
of cohomological Yang-Mills
theory~\cite{Hirano:1997ai,Moore:1998et,Kazakov:1998ji}, we have not
succeeded in finding analogous twists for our reduced Chern-Simons
quiver models. Nevertheless, by using the duality with the IKKT model
we are able to construct a cohomological
matrix model with $\CN=2$ supersymmetry ``by hand'' which possesses
the desired properties and should capture the salient features of the
index we are after. The major difference in our treatment
is that we regard the F-term relations for the bifundamental component scalars
as relations in the double of the quiver characterising the
Chern-Simons-matter theory, as opposed to the actual relations of the
quiver which are the F-term relations of the supersymmetric gauge
theory among chiral superfields derived from the superpotential; this
means that our moduli space is built on the cotangent bundle of the
representation variety of the original quiver with fixed dimension vector.
For the reduced ABJM theory, this quiver is the well-studied
conifold quiver and we are able to take
this calculation through to the end. We explicitly compute an
equivariant index which enumerates $\sSU(4)$ R-charge assignments of
framed BPS states; the equivariant index receives contributions from
only those BPS states which are fixed by the action of the maximal
torus of the R-symmetry group, and its explicit combinatorial formula is given as a sum over
filtered pyramid partitions related to length two empty room configurations. This index can
also be used to compute a regularised volume of the non-compact vacuum
moduli space. We suggest that
this quantity can be interpreted as a character in a dimensional
reduction of M-theory to one dimension; however, a complete
physical interpretation of our equivariant index is not clear at
the moment, and in particular if it can be interpreted in some way as a
``superconformal index''.

The structure of the remainder of this paper is as follows. In
\S\ref{ternary} we review some ternary algebra structures, in
particular 3-Lie algebras, hermitian 3-algebras, and some specific
examples we use later in the paper. \S\ref{dimredCS} is dedicated to
constructing the various reduced Chern-Simons quiver models we
study. In \S\ref{3algebramodels} we explain how these quiver models
are related to the reduced 3-algebra model introduced
in~\cite{DeBellis:2010sy} by taking various scaling limits, or by
making particular choices of 3-algebra. The duality between the
Chern-Simons quiver matrix models and the IKKT matrix model using the
Mukhi-Papageorgakis map~\cite{Mukhi:2008ux} is explained in
\S\ref{IKKTmodels} In \S\ref{twistedBLG} we apply the
Mukhi-Papageorgakis map to a particular topological twist of the BLG
theory, thus yielding a topologically twisted version of
$\mathcal{N}=8$ supersymmetric Yang-Mills theory in three
dimensions; we dimensionally reduce this theory to find a new
topologically twisted version of
the IKKT matrix model, and explain why it is not possible to lift this
new twist to the reduced ABJM theory. Finally, in
\S\ref{equiv3algebra} we construct a cohomological matrix model
related to the reduced ABJM theory and compute an equivariant index using localization methods.

\section{Ternary algebras\label{ternary}}

In this section we review the definitions and properties of the
3-algebras that we will encounter in this paper, primarily to set up
our conventions and notation. 

\subsection{Metric 3-Lie algebras}

A metric 3-Lie algebra is a finite-dimensional real vector space $\mathcal{A}$ equipped with a
positive-definite symmetric bilinear form $(-,-):\CA\times\CA\to\FR$
and a totally antisymmetric trilinear map $[-,-,-] :\mathcal{A}\times
\CA\times\CA \rightarrow \mathcal{A}$. The 3-bracket satisfies the {fundamental identity}
\begin{align}
[X,Y,[Z_1,Z_2,Z_3]]= [[X,Y,Z_1],Z_2,Z_3]+[Z_1,[X,Y,Z_2],Z_3] +[Z_1,Z_2,[X,Y,Z_3]]
\label{metricfundid}\end{align}
for $X,Y,Z_1,Z_2,Z_3 \in\CA$, and it is compatible with the metric in
the sense that
 \begin{align}
 ( [Z_1,Z_2,X],Y) + (X, [Z_1,Z_2,Y]) = 0 ~~. 
\label{metricity} \end{align}
Given a basis $\{\tau_a\}$ of generators for $\CA$, the 3-bracket
\begin{align}
[\tau_a,\tau_b,\tau_c]=f_{abcd}\,\tau_d
\end{align}
gives the totally antisymmetric structure constants $f_{abcd}$ of
$\CA$. 

Every metric 3-Lie algebra admits an {associated Lie algebra},
denoted $\mathfrak{g}_{\mathcal{A}}$.  Given
$(Y_1,Y_2)\in\CA\wedge\CA$, we define an operator
$D_{(Y_1,Y_2)}\in\sEnd(\CA)$ by the 3-Lie bracket
\begin{align}
D_{(Y_1,Y_2)}(X) = [Y_1,Y_2,X] \ .
\end{align}
The fundamental identity (\ref{metricfundid}) implies that $D_{(Y_1,Y_2)}$ is an
inner derivation of $\CA$. The linear span of the collection of operators $\{D_{(Y_1,Y_2)}\}$ forms the
Lie algebra $\frg_\CA$ which is a Lie subalgebra of $\aso(\CA)$: The fundamental identity guarantees closure of
the commutator bracket in
$\sEnd(\CA)$ which is expressed in terms of the 3-Lie bracket as
\begin{align}
[D_{(Y_1,Y_2)},D_{(Z_1,Z_2)}](X) =
[[Y_1,Y_2,Z_1],Z_2,X]+[Z_1,[Y_1,Y_2,Z_2],X ] \ .
\end{align}
The Jacobi identity for $\frg_\CA$ also follows from the fundamental
identity (\ref{metricfundid}), while the compatibility condition (\ref{metricity}) implies that the
metric of $\CA$ is $\frg_\CA$-invariant; moreover, the metric on $\CA$
induces an ad-invariant symmetric bilinear form
$\Tr_{\frg_\CA}$ on
$\frg_\CA$ given by
\beq
\Tr_{\frg_\CA}\big(D_{(Y_1,Y_2)}\,D_{(Z_1,Z_2)}\big):=
\big([Y_1,Y_2,Z_1]\,,\, Z_2\big) \ ,
\label{TrgA}\eeq
for which every element is null,
i.e. $\Tr_{\frg_\CA}\big(D_{(Y_1,Y_2)}^2\big)=0$.
The generators of $\mathfrak{g}_{\mathcal{A}}$ are the operators $D_{ab}\in\sEnd(\CA)$ given by
 \begin{align}
 D_{ab}(\tau_c)= f_{abcd}\, \tau_d ~~.
 \end{align}
Together with a real finite-dimensional orthogonal representation, the
subalgebra $\frg_\CA$ can be used to reconstruct the 3-Lie algebra
$\CA$ via the Faulkner construction~\cite{deMedeiros:2008zh}.
 
 One can also reduce a 3-Lie algebra $\CA$ to a Lie algebra $\CA'$ which is
 generally different from $\mathfrak{g}_{\mathcal{A}}$~\cite{Filippov:1985aa}.  One chooses a fixed element $Z_0 \in \mathcal{A}$ and identifies the vector space $\mathcal{A}^{\prime}$ with $\mathcal{A}$. The Lie bracket on $\mathcal{A}^{\prime}$ is defined as
 \begin{align}\label{reducedbracket}
 [X,Y]=[X,Y,Z_0]
 \end{align}
 for $X,Y\in\CA'$, and the corresponding Jacobi identity for $\CA'$
 likewise follows from (\ref{metricfundid}).

All 3-Lie algebras with metric of euclidean signature are direct sums of the
four-dimensional 3-algebra $\CA=A_4$. It is defined by generators $\tau_a$,
$a=1,2,3,4$, obeying the relations
$[\tau_a,\tau_b,\tau_c]=\epsilon_{abcd}\, \tau_d$ and $(\tau_a,\tau_b)=\delta_{ab}$. The associated Lie
algebra is $\frg_{A_4}=\aso(4)=\aso(3)\oplus \aso(3)$, while the
reduced Lie algebra for any fixed $Z_0=\tau_a$ is $A_4'=
\aso(3)=\asu(2)$. On the other hand, applying the Faulkner
construction to the pair $(\aso(4),\FR^4)$ with inner product on
$\aso(4)$ given by the Cartan-Killing form makes the fundamental
representation $\FR^4$ into a Lie triple system~\cite{deMedeiros:2008zh}.
 
\subsection{Hermitian 3-algebras}

We will now relax the requirement of total
antisymmetry of the 3-bracket; these 3-algebras are generally called 3-Leibniz
algebras. Here we are interested in the special class of 3-Leibniz
algebras called hermitian 3-algebras. They comprise a complex metric
3-algebra which is a finite-dimensional complex vector space
$\mathcal{A}$ equipped with a
hermitian inner product $(-,-)$ and a trilinear map
$[-,-;-]:\CA\times\CA\times\CA \to \CA$. We require that the 3-bracket is
antisymmetric in its first two entries only, and that it is complex linear in
its first two arguments and complex antilinear in its third argument.
It satisfies a version of the {fundamental identity} given by
\begin{align}\label{complexfundamentalidentity}
[[Z_1
,Z_2
;Z_3
],X
;Y
]= [[Z_1
,X
;Y
],Z_2
;Z_3
]+[Z_1
,[Z_2
,X
;Y
],Z_3
] - [ Z_1
,Z_2
;[Z_3
,Y
;X
]] \ ,
\end{align}
and also the metric {compatibility conditions}
 \begin{align}\label{compatibility}
 ( [Z_1
,Z_2
;X
],Y
)- ([Z_2
,Z_1
;Y
],X
) =&\ 0 \ , \nonumber \\[4pt]
 ( [X
,Z_1
;Z_2
],Y
) - (X
, [Y
,Z_2
;Z_1
]) =&\ 0 ~.
 \end{align}
On generators $\{\tau_a\}$ for $\CA$ the 3-bracket
\begin{align}
[\tau_a
,\tau_b
;\tau_c
]=f_{abcd}\, \tau_d
\end{align} 
defines the structure constants $f_{abcd}$ of $\CA$; they are antisymmetric in
their first two indices and have the additional symmetry properties
$f_{abcd}=f_{cdab}=f_{badc}$.

Every complex metric 3-algebra satisfying the fundamental identity
(\ref{complexfundamentalidentity}) admits an {associated Lie algebra} $\mathfrak{g}_{\mathcal{A}}$. The generators of $\mathfrak{g}_{\mathcal{A}}$ are defined to be operators $D_{ab}\in\sEnd(\CA)$ expressed in terms of the 3-bracket as
 \begin{align}
 D_{ab}(\tau_c):=[\tau_c
,\tau_a
;\tau_b
] = f_{cabd}\, \tau_d ~. 
 \end{align}
The fundamental identity guarantees closure of their commutator bracket
 \begin{align}
 [D_{ab},D_{cd}](\tau_e)=[[\tau_e
,\tau_c
;\tau_d
],\tau_a
;\tau_b
]-[[\tau_e
,\tau_a
;\tau_b
],\tau_c
;\tau_d
]~,
 \end{align}
as well as the Jacobi identity for $\frg_\CA$.
 
In this paper we are primarily interested in the following hermitian
3-algebra. Consider the vector space $\mathcal{A}=\sHom(V_L,V_R)$ of
linear maps $X: V_L \rightarrow V_R$ between two complex inner product spaces
$V_L$ and $V_R$. The 3-bracket defined by
\begin{align}\label{hermitianbracket}
[X,Y;Z]=\lambda\, (X\,Z^\dag\, Y-Y\,Z^\dag\, X) \ ,
\end{align}
for an arbitrary constant $\lambda\in\FC$, satisfies the
fundamental identity \eqref{complexfundamentalidentity}. The metric on $\CA$ given
by the Schmidt inner product
\begin{align}\label{hermitianinnerproduct}
(X,Y)=\Tr_{V_L}(X^\dag\, Y)
\end{align}
then satisfies the compatibility conditions
\eqref{compatibility}. This 3-algebra has associated
Lie algebra $\frg_\CA= \mathfrak{u}(V_L)\oplus\mathfrak{u}(V_R)$: An
endomorphism $\phi=(\phi_L,\phi_R)\in\frg_\CA$ acts on $X\in\CA$ as
\beq
\phi X=X\, \phi_L-\phi_R\, X \ .
\eeq

\subsection{Lorentzian 3-Lie algebras}

A large class of 3-Lie algebras $\mathcal{A}_{\frh}$ with compatible
metric of lorentzian signature are described as the semisimple
indecomposable lorentzian 3-Lie algebras of dimension $d+2$ which are obtained
by double extension from a semisimple Lie algebra
$\mathfrak{\frh}$ of dimension $d$~\cite{DeMedeiros:2008zm}. Let $\tau_a$, $a=1,\dots,d$, be a set of generators for $\frh$ with
antisymmetric structure constants $f_{abc}$ defined by the Lie bracket
$[\tau_a,\tau_b]=f_{abc}\, \tau_c$. The 3-Lie algebra $\CA_\frh$ has generators
$\tau_a$, $\tau_0$ and $\unit$ with the 3-bracket relations
\begin{align}\label{nappiwitten3liealgebra}
\begin{tabular}{lll}
$[\tau_a,\tau_b,\tau_c]=f_{abc}\, \unit~~,$&$[\tau_0 ,\tau_a,\tau_b]=f_{abc}\, \tau_c~~,$& 
$[\unit,\tau_a,\tau_b]=0 = [\unit,\tau_a,\tau_0 ]$ 
\end{tabular}
\end{align}
and the inner product relations
\begin{align}\label{nappiwitteninnerproduct}
\begin{tabular}{lll}
$(\unit,\unit)=0~,$&$(\unit,\tau_a
)=0~,$&$(\unit,  \tau_0)=-1~,$ \\[4pt]
$(\tau_0,\tau_a)=0~,$&$ (\tau_0,\tau_0)=\beta~,$&$(\tau_a,\tau_b)=\delta_{ab}~,$
\end{tabular}
\end{align}
where $\beta\in\FR$ is an arbitrary constant. Note that with $Z_0=\tau_0$, the reduced bracket (\ref{reducedbracket}) coincides with the
  Lie bracket of $\frh$ and $\CA_\frh'=\frh\oplus\FR$. On the other hand, the associated Lie algebra of
  $\CA_\frh$ is the semi-direct sum
\begin{align}
\mathfrak{g}_{\mathcal{A}_{\frh}} = \fru(1)^d\ltimes \frh \ .
\end{align}

\section{Dimensional reduction of Chern-Simons-matter theories\label{dimredCS}} 

In
this section we study the dimensional reduction of three-dimensional
supersymmetric Chern-Simons-matter theories to
zero dimensions. The resulting reduced models are supersymmetric quiver matrix
models which will be the focal points
of our analysis in this paper. Depending on the choice of gauge
group and superpotential, one is able to construct models with varying
amounts of supersymmetry. From the point of view of the underlying
quivers, the quiver theories we construct are quiver quantum mechanics
describing BPS particles; the representations of the quiver then
correspond to BPS bound states. Although the Chern-Simons level $k$ must be an integer to ensure large
gauge invariance in the field theory, our matrix integrals are well-defined for
non-integer $k$ and hence we do not impose any quantization condition
on the Chern-Simons coupling constant in what follows.

\subsection{$\mathcal{N}=2$ Chern-Simons quiver matrix models}

The field content for the $\mathcal{N}=2$ supersymmetric Chern-Simons
gauge multiplet $\mbf V$ in three-dimensional flat space $\FR^{1,2}$
consists of a gauge field $A_\mu$, $\mu=0,1,2$,
two auxiliary scalar fields $D$ and $\sigma$, and a two-component complex auxiliary fermion field $\lambda$.  The fields are
valued in the Lie algebra $\mathfrak{g}$ of a matrix gauge group $G$.
The action is given by
\begin{align}\label{generalchernsimons}
S_{\rm g}=\int\, \dd^3x\ \kappa\,\Tr_\frg\Big(\epsilon^{\mu\nu\lambda}\,
\big(A_\mu \, \partial_\nu A_\lambda+\tfrac{2\, \di}{3}\, A_\mu \, A_\nu
\, A_\lambda\big)-\bar{\lambda}\, \lambda+2D\, \sigma\Big)~,
\end{align}
where $\kappa\in\FR$ is a coupling constant and $\Tr_\frg$ is an invariant quadratic
form on the Lie algebra $\frg$. The generators of the Clifford algebra
$C\ell(\mathbb{R}^{1,2})$ are the gamma-matrices
$\gamma^\mu$ which satisfy the anticommutation relations
\beq
\big\{\gamma^\mu,\gamma^\nu\big\}=2\eta^{\mu\nu}
\eeq
and are taken to be Pauli spin matrices
\beq
\gamma^0=\begin{pmatrix} 1&0\\0&-1 \end{pmatrix} \ , \qquad
\gamma^1=\begin{pmatrix} 0&1\\1&0 \end{pmatrix} \ , \qquad
\gamma^2= \begin{pmatrix} 0&-\di \\ \di & 0\end{pmatrix} \ , 
\eeq
while the spinor adjoint is
\begin{align}
\bar{\lambda}=\lambda^\dag\, \gamma^0~.
\end{align}
We perform a dimensional reduction to zero dimensions in which the
gauge fields $A_\mu$ become a collection of $\frg$-valued scalar fields, and
similarly for the other fields of $\mbf V$.  The reduced action is
\begin{align}
\mathcal{S}_{\rm g}=\kappa\,\Tr_\frg\big(\tfrac{2\,\di}{3}\,
\epsilon^{\mu\nu\lambda}\, A_\mu \, A_\nu \, A_\lambda-\bar{\lambda}\, \lambda+2D\,
\sigma\big) ~.
\end{align}
This action is invariant under the $\CN=2$ supersymmetry transformations
\begin{align}
\delta
A_\mu=&\ \tfrac{\di}{2}\, (\bar{\eta}\, \gamma_\mu\, \lambda-\bar{\lambda}\,
\gamma_\mu\, \eps)~,
\nonumber \\[4pt]\nonumber
\delta\sigma=&\ \tfrac{\di}{2}\, (\bar{\eta}\, \lambda-\bar{\lambda}\,
\eps)~, \\[4pt] \nonumber
\delta D=&\ \tfrac{\di}{2}\, (\bar{\eta}\, \gamma^\mu\,
[A_\mu,\lambda]+[A_\mu,\bar{\lambda}]\, \gamma^\mu\,
\eps)+\tfrac{\di}{2}\, (\bar{\eta}\,
[\lambda,\sigma]+[\bar{\lambda},\sigma]\, \eps)~, \\[4pt] \nonumber
\delta\lambda=& \ -\di\, \big(\tfrac{1}{2}\, \gamma^{\mu\nu}\, [A_\mu,A_\nu]+
D+\gamma^\mu\, [A_\mu,\sigma] \big)\, \eps~, \\[4pt] 
\delta\bar\lambda=& \ \di\, \bar\eta\, \big(- \tfrac{1}{2}\, \gamma^{\mu\nu}\, [A_\mu,A_\nu]+
D+\gamma^\mu\, [A_\mu,\sigma] \big)~, \label{genN=2susy}
\end{align}
where ${\eta}$ and $\eps$ are two
independent Dirac spinors of $\sSO(1,2)$ and
$\gamma^{\mu\nu}:= \frac12\, [\gamma^\mu,\gamma^\nu]$.
The two supersymmetry transformations generated by ${\eta}$ or $\eps$
alone commute. The commutator of an ${\eta}$-supersymmetry
with an $\eps$-supersymmetry generates a sum of a gauge
transformation, a Lorentz rotation, a dilatation, and an R-rotation.

This action can be extended to include supersymmetric matter fields. The
matter content is a chiral multiplet $\mbf\Phi$ with
component fields $\mbf\Phi=(
Z, Z^\dag, \psi, \bar{\psi}, F, F^\dag)
$, which are also valued in the Lie algebra $\frg$. The field $Z$ is a complex
matter field, $F$ is an auxiliary complex scalar field, and $\psi$ is
a two-component Dirac spinor field.  The action reads
\begin{align}
S_{\rm m}=& \ \int\,\dd^3x \ \Tr_\frg\big(\nabla_\mu Z^\dag \, \nabla^\mu
Z-Z^\dag\, \sigma^2\, Z+Z^\dag\, D\, Z+F^\dag\, F \nonumber \\ &
\qquad \qquad \qquad +\di\, \bar{\psi}\,
\gamma^\mu\, \nabla_\mu \psi-\bar{\psi}\, \sigma\, \psi-\di\,
\bar{\psi}\, \lambda\, Z+\di\, Z^\dag\, \bar{\lambda}\, \psi\big)~,
\end{align}
where the gauge covariant derivatives act as $\nabla_\mu
Z:= \partial_\mu Z+\di\, [A_\mu,Z]$.
We perform a dimensional reduction as above, so that the reduced matter action reads as
\begin{align}
\mathcal{S}_{\rm m}=&\ \Tr_\frg\big( -[A_\mu,Z^\dag]\, [A^\mu,Z]-Z^\dag\,
\sigma^2\, Z+Z^\dag\, D \, Z+F^\dag \, F \nonumber\\[4pt] & \qquad
\qquad -\bar{\psi}\, \gamma^\mu\,
[A_\mu,\psi]-\bar{\psi}\, \sigma\, \psi-\di\, \bar{\psi}\, \lambda\,
Z+\di\, Z^\dag\, \bar{\lambda}\, \psi\big) ~.
\label{CSquivermatter}\end{align}
The supersymmetry transformations are given by
\begin{align}
\delta Z=&\ \bar{\eta}\, \psi~, \nonumber \\[4pt]
\delta Z^\dag=&\ \bar{\psi} \, \eps~, \nonumber \\[4pt]
\delta\psi=&\ \di\, \big(\gamma^\mu\, [A_\mu,Z]-\sigma\, Z\big)\, \eps+F\, \eps^\ast~, \nonumber \\[4pt]
\delta\bar{\psi}=&\ \di\, \bar{\eta}\, \big(\gamma^\mu\, [A_\mu, Z^\dag]+ Z^\dag\,
\sigma\big) ~, \nonumber \\[4pt]
\delta F=&\ \bar{\eta}\,^\ast\,\big(\gamma^\mu\,[A_\mu,\psi]+\di\, \lambda\, Z+\sigma\, \psi\big)~. 
\end{align}

The complete action of the reduced $\CN=2$ Chern-Simons-matter theory thus reads as
\begin{align}\label{reducedgeneralchernsimons}
\mathcal{S}=&\ \Tr_\frg\Big(\kappa\, \big(\tfrac{2\, \di}{3}\,
\epsilon^{\mu\nu\lambda}\, A_\mu \, A_\nu \, A_\lambda-\bar{\lambda}\, \lambda+2D\, \sigma
\big) -[A_\mu,Z^\dag]\, [A^\mu,Z]-Z^\dag\, \sigma^2\, Z+Z^\dag\, D \,
Z+F^\dag\, F \nonumber \\[4pt]
& \qquad \qquad -\bar{\psi}\, \gamma^\mu\, [A_\mu,\psi]-\bar{\psi}\, \sigma\, \psi-\di\,
\bar{\psi}\, \lambda \, Z+\di \, Z^\dag\, \bar{\lambda}\, \psi\Big) ~.
\end{align}
The BRST transformations imply that the supersymmetric configurations
satisfy
\beq
[A_\mu,A_\nu]=0=[A_\mu,\sigma] \ , \qquad [A_\mu,Z]=0=[A_\mu,Z^\dag] \
, \qquad D=0=F \ .
\label{N=2BPS}\eeq

When the gauge group is a product of unitary groups
\beq
G=\prod_{a=1}^r\, \sU(N_a) \ ,
\label{Gprod}\eeq
we decompose the
reduced vector multiplet as $\mbf V=\bigoplus_{a}\, \mbf V^a$ where
$\mbf V^a\in\sEnd(V_a)$ are regarded as linear transformations of
complex inner product spaces $V_a=\FC^{N_a}$ for $a=1,\dots,r$, while the reduced matter
multiplet is decomposed as $\mbf\Phi=\bigoplus_{a,b}\, \mbf\Phi^{ab}$
with $\mbf\Phi^{ab}\in\sHom(V_a,V_b)$ and
$\mbf\Phi_{ab}^\dag\in\sHom(V_b,V_a)$ for $a,b=1,\dots,r$; then
$\Tr_\frg$ refers to the trace in the fundamental representation of
$G$ which is possibly graded
over the factors of $G$. In this case the supersymmetric Chern-Simons-matter theory
reduces to a quiver matrix model, which defines a
finite-dimensional representation of the double of the quiver with $r$ nodes that carry the
gauge degrees of freedom $A_\mu^a$ (plus their superpartners and
auxiliary fields) transforming in the adjoint representation of $\sU(N_a)$, and with an arrow from node $a$ to node $b$ for
every non-zero matter field $Z^{ab}$ (plus their superpartners and
auxiliary fields) transforming in the bifundamental representation
of $\sU(N_a)\times\sU(N_b)$, along with an arrow in the opposite direction for
the adjoint $Z_{ab}^\dag$. The double quiver is further equiped with a
set of
relations among the arrows that follow from the BPS equations of the
supersymmetric gauge theory, which define a system of static quiver
vortices; geometrically, representations of the double quiver are
cotangent to representations of the original quiver. In this paper we will use this double quiver to construct
and study the vacuum moduli spaces of supersymmetric
Chern-Simons-matter theories in three dimensions in terms of moduli
spaces of quiver representations.

\subsection{$A_1$ quiver matrix model}

The simplest example of the above construction is with a product
gauge group
\beq
G={\sf U}(N_L) \times {\sf U}(N_R) \ .
\eeq
The matter content $\mbf\Phi$
of the theory provides a representation of the double of the $A_1$ quiver
\beq
\xymatrix@C=20mm{
\bullet \ar[r] & \bullet
}
\label{A1quiver}\eeq
We place complex inner product spaces $V_L=\FC^{N_L}$ and
$V_R=\FC^{N_R}$ at the left and right nodes of the quiver
(\ref{A1quiver}), respectively.
The matter field is regarded as a linear map $Z:V_L\rightarrow V_R$
representing the arrow of the quiver (\ref{A1quiver}), with hermitian conjugate $Z^\dag:V_R\rightarrow V_L$. 
The matrices $Z$, $F$ and $\psi$ are bifundamental fields, i.e. they
transform in the fundamental representation of ${\sf U}(N_R)$ and in the anti-fundamental representation of ${\sf U}(N_L)$.
The vector multiplet has field content
$\mbf
V=\big(A^{L,R}_\mu,\sigma^{L,R},\lambda^{L,R},\bar{\lambda}^{L,R},D^{L,R} \big)$.
The matrices $A^{L,R}_\mu\in\sEnd(V_{L,R})$ for $\mu=0,1,2$ transform in
the adjoint representation of $\sU(N_{L,R})$, $\lambda^{L,R}$ are two-component complex fermionic matrices, while $\sigma^{L,R}$ and
$D^{L,R}$ are auxiliary matrix fields.
The invariant quadratic form is given by
$\Tr_\frg=\Tr_{V_L}\oplus(-\Tr_{V_R})$, and the action of the
quiver matrix
model takes the form
\begin{align}\label{bifundamentalmatter}
\mathcal{S}_{A_1}=&\, \Tr_V\bigg(\kappa\,\Big( \tfrac{2\,\di}{3}\,
\epsilon^{\mu\nu\lambda}\big(A^L_\mu \, A^L_\nu \, A^L_\lambda-A^R_\mu \, A^R_\nu \, A^R_\lambda
\big)-\bar{\lambda}^L\, \lambda^L+\bar{\lambda}^R\, \lambda^R+2D^L\,
\sigma^L-2D^R\, \sigma^R \Big) \nonumber \\ \nonumber
&\qquad \qquad -\big(A^L_\mu\, Z^\dag-Z^\dag\, A^R_\mu\big)\, \big(A^R_\mu\, Z-Z\,
A^L_\mu\big)-\bar{\psi}\, \gamma^\mu\, \big(A^R_\mu\, \psi-\psi\, A^L_\mu
\big)+F^\dag \,F\\ \nonumber & \qquad \qquad +Z\, D^L\, Z^\dag
-Z^\dag\, D^R \, Z -\di\,\bar{\psi}\, Z\, \lambda^L +\di\, \bar{\psi}\,
\lambda^R\, Z+\di \, \bar{\lambda}^L\, Z^\dag \, \psi-\di \, Z^\dag\,
\bar{\lambda}^R\, \psi \\ & \qquad \qquad +Z^\dag\, Z\, \sigma^2_L -Z^\dag \,
\sigma^2_R \, Z +2Z^\dag\, \sigma^R\,  Z\, \sigma^L
-\bar{\psi}\, \psi\, \sigma^L +\bar{\psi}\, \sigma^R\, \psi\bigg)~,
\end{align}
where the trace is taken over $V=V_L$ or $V=V_R$ where appropriate.
The supersymmetry transformations of this matrix model are
given by
\begin{align} \nonumber
\delta A^{L,R}_\mu =&\ \tfrac{\di}{2}\,\big(\bar{\eta}\, \gamma_\mu
\, \lambda^{L,R}-\bar{\lambda}^{L,R}\, \gamma_\mu \, \eps\big)~,\\[4pt] \nonumber
\delta \sigma^{L,R}=&\ \tfrac{\di}{2}\, \big(\bar{\eta}\,
\lambda^{L,R}-\bar{\lambda}^{L,R}\, \eps\big)~, \\[4pt] \nonumber
\delta D^{L,R}=&\ \tfrac{\di}{2}\, \bar{\eta}\, \gamma^\mu \,
\big[A^{L,R}_\mu ,\lambda^{L,R}\big]+\tfrac{\di}{2}\, \big[A^{L,R}_\mu
,\bar{\lambda}^{L,R}\big]\, \gamma^\mu \, \eps +\tfrac{\di}{2}\,
\bar{\eta}\, \big[\lambda^{L,R},\sigma^{L,R}\big]+\tfrac{\di}{2}\,
\big[\bar{\lambda}^{L,R},\sigma^{L,R}\big]\, \eps~, \\[4pt] \nonumber
\delta \lambda^{L,R}=&\ \di\,\big(\tfrac{1}{2}\, \gamma^{\mu\nu}\,
\big[A^{L,R}_\mu ,A^{L,R}_\nu\big]-D^{L,R}-\gamma^\mu \,
\big[A^{L,R}_\mu ,\sigma^{L,R}\big]\big)\, \eps~, \\[4pt]\nonumber
\delta Z=&\ \bar{\eta}\, \psi~, \\[4pt] \nonumber
\delta Z^\dag=&\ \bar{\psi}\, \eps~, \\[4pt] \nonumber
\delta\psi=&\ \di\, \gamma^\mu \, \big(Z\, A^L_\mu - A^R_\mu \,Z 
\big)\, \eps-\di\, \eps\, \big(Z\, \sigma^L-\sigma^R \,Z \big)+F\,
\eps^\ast~, \\[4pt] \nonumber
\delta\bar\psi = & \ \di\, \bar\eta \, \gamma^\mu\, \big(Z^\dag \, A^R_\mu-
A_\mu^L \, Z^\dag\big) +\di\, \bar\eta\,
\big(\sigma^L \,Z^\dag-Z^\dag \,
\sigma^R\big) \ , \\[4pt] 
\delta F=& \ \bar\eta\,^\ast\,\Big(\gamma^\mu \, \big(\psi \,
A^L_\mu-A^R_\mu\, \psi \big)+\di\, \big(Z\, \lambda^L-\lambda^R \,
Z\big)+\big(\psi\, \sigma^L-\sigma^R \, \psi\big)\Big)~. \label{A1susy}
\end{align}

It is straightforward to extend this construction to give a 
quiver matrix model based on the $A_{r-1}$ Dynkin diagram for all
$r\geq2$; the corresponding linear $A_{r-1}$ quiver is a chain of $r$ nodes
with $r-1$ arrows connecting nearest neighbour vertices.

\subsection{ABJM matrix model\label{ABJMmatrix}}

The ABJM
model~\cite{Aharony:2008ug,Aharony:2008gk,VanRaamsdonk:2008ft} 
is based on adding arrows to the $A_1$ quiver to give the ABJM quiver

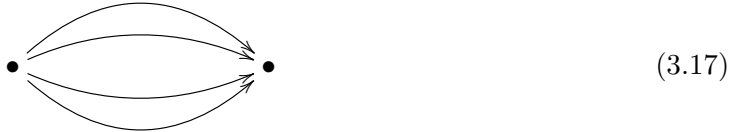
\begin{equation}
\vspace{10pt}
\begin{xy}
\xymatrix@C=30mm{
\bullet \ar@/^1pc/[r] \ar@/^2pc/[r]  \ar@/_1pc/[r]  \ar@/_2pc/[r] & \bullet
}
\end{xy}
\vspace{10pt}
\label{quiverC3Z3}\end{equation}
There are now four complex bifundamental matter fields $\mbf\Phi^i=(Z^i,F^i,\psi^i)$, which further transform under the
R-symmetry group ${\sf SU}(4)$ which acts as rotations of the flavour
index $i=1,2,3,4$. We deform the generic $\CN=2$ supersymmetric
Chern-Simons quiver matrix model (\ref{reducedgeneralchernsimons}) by
adding a suitable quartic superpotential of the chiral
superfields $\mbf\Phi^i$~\cite{Gaiotto:2007qi,Benna:2008zy} which reads as
\beq
\CW(\mbf\Phi) = \tfrac{\kappa}{4!} \, \epsilon^{ijkl}\,
\Tr_V\big(\mbf\Phi_i\, \mbf\Phi_j^\dag\, \mbf\Phi_k\, \mbf\Phi_l^\dag \big) \ .
\label{ABJMsuperpot}\eeq
The extrema of the superpotential define the relations of the double quiver associated to
the ABJM quiver \eqref{quiverC3Z3}. After integrating out the
auxiliary fields, the action is then
\begin{align}\nonumber
\mathcal{S}_{\rm ABJM}=&\ \Tr_V\Big(\tfrac{2\,\di}{3}\, \kappa\,
\epsilon^{\mu\nu\lambda}\, \big(A^L_\mu \, A^L_\nu \, A^L_\lambda-A^R_\mu \,
A^R_\nu \, A^R_\lambda\big) \\ \nonumber
& \qquad -2A^L_\mu\, Z^\dag_i\, A^R_\mu \, Z^i+A^L_\mu\,  Z^\dag_i\,
Z^i\, A^L_\mu +A^R_\mu \, Z^i\, Z^\dag_i\, A^R_\mu
-\di\,\bar{\psi}_i\, \gamma^\mu \, A^R_\mu \, \psi^i+\di\, \bar{\psi}
_i\, \gamma^\mu \, \psi^i\, A^L_\mu \\ \nonumber
& \qquad +\tfrac{\di}{2\kappa}\, \big(Z^\dag_i\, Z^i\, \bar{\psi}
_j\, \psi^j-\bar{\psi}
_j\, Z^i\, Z^\dag_i\, \psi^j-2Z^\dag_i\, Z^j\, \bar{\psi}
^i\, \psi_j \\ \nonumber
&\qquad \qquad \qquad +2\bar{\psi}
^j\, Z^i\, Z^\dag_j\, \psi_i -\epsilon^{ijkl}\, Z^\dag_i\, \psi_j  \,
Z^\dag_k\, \psi_l+\epsilon_{ijkl}\, Z^i\, \bar{\psi}
^j\, Z^k\, \bar{\psi}
^l \big) \\ \nonumber
&\qquad +\tfrac{1}{12\kappa^2}\, \big(Z^i\, Z^\dag_i\, Z^j\,
Z^\dag_j\, Z^k\, Z^\dag_k+Z^\dag_i\, Z^i\, Z^\dag_j \, Z^j\,
Z^\dag_k\, Z^k \\ 
& \qquad \qquad \qquad +4Z^i\, Z^\dag_j\, Z^k\, Z^\dag_i\, Z^j\,
Z^\dag_k-6Z^i\, Z^\dag_j\, Z^j\, Z^\dag_i\, Z^k\, Z^\dag_k\big)\Big)~.
\label{reducedabjm} \end{align}
The corresponding supersymmetry transformations are
\begin{align}
\delta Z^i &= \di \, \omega^{ij} \, \psi_j~, \nonumber \\[4pt]
\delta Z^\dag_i &= \di \, \psi^\dag_{j} \, \omega_{ij}~, \nonumber \\[4pt]
\delta \psi_i &= - \gamma^\mu  \,\omega_{ij} \,\big(Z^j \,
A^L_\mu-A^R_\mu \,Z^j \big)-\tfrac{1}{2\kappa} \,\Big( \omega_{ij} \,
\big(Z^k\, Z^\dag_k \, Z^j-Z^j\, Z^\dag_k\, Z^k \big)-2 \omega_{kl}
\,Z^k \, Z^\dag_i\, Z^l\Big)~, \nonumber \\[4pt]
\delta \bar{\psi}{}^{i} &= \big(Z^\dag_j \, A^R_\mu- A^L_\mu \,Z^\dag_j
\big) \, \omega^{ij} \, \gamma^\mu  - \tfrac{1}{2\kappa} \, \Big(
\big(Z^\dag_j\, Z^k\, Z^\dag_k -Z^\dag_k\, Z^k\, Z^\dag_j \big) \,
\omega^{ij} - 2 Z^\dag_l\, Z^i\, Z^\dag_k\, \omega^{kl} \Big)~,
\nonumber \\[4pt]
\delta A^L_\mu  &= -\tfrac{\di}{4\kappa}\, \big(Z^i \,\psi^\dag_{j}
\,\gamma_\mu \, \omega_{ij} - \omega^{ij} \, \gamma_\mu \, \psi_i \,
Z^\dag_j\big)~, \nonumber \\[4pt]
\delta A^R_\mu  &= -\tfrac{\di}{4\kappa}\, \big(\psi^\dag_{i}\, Z^j \,
\gamma_\mu \, \omega_{ij} - \omega^{ij} \, \gamma_\mu \, Z^\dag_i \, \psi_j\big)~,
\label{ABJMsusy}\end{align}
where $\omega^{ij}$ are $\CN=6$ supersymmetry transformation
parameters obeying $\omega^{ij}=(\omega_{ij})^*=-\frac12\,
\epsilon^{ijkl}\, \omega_{kl}$. 

The BPS equations of the ABJM theory were derived in
\cite{Kim:2009ny}; here we present them for the dimensionally reduced
model. They are determined by the quantities
\begin{align}\label{abjmbeta}
\CZ^{jk}_i:= Z^j\, Z^\dag_i\, Z^k-Z^k\, Z_i^\dag\, Z^j
\end{align}
for $j<k$.
We set the fermions equal to zero. The BPS equations for the
supersymmetric solutions of the matrix model then follow from
the fermionic supersymmetry variations in (\ref{ABJMsusy}) using the
independence of the gamma-matrices as a basis of the Clifford algebra, and are
given by
\begin{align} \nonumber
\big[A_\mu^L,A_\nu^L\big]&=0 = \big[A_\mu^R,A_\nu^R\big]~, \\[4pt] \nonumber
A^L_1\,Z^1-Z^1\, A^R_1-\di\,\big(A^L_2 \, Z^1-Z^1\, A^R_2\big)&=0~, \\[4pt] \nonumber
A^L_\mu \, Z^i-Z^i\, A^R_\mu &=0~~ (i\neq 1~,~ \mu=1,2)~, \\[4pt] \nonumber
A^L_0\, Z^2-Z^2\, A^R_0-\di\,\CZ^{21}_1&=0~, \\[4pt] \nonumber
A^L_0\, Z^3-Z^3\, A^R_0-\di\, \CZ^{13}_1&=0~, \\[4pt] \nonumber
A^L_0\, Z^4-Z^4\, A^R_0-\di\, \CZ^{14}_1&=0~, \\[4pt] \nonumber
\CZ^{31}_3=\CZ^{41}_4&=\CZ^{21}_3~, \\[4pt] \nonumber
\CZ^{43}_4=\CZ^{34}_3=\CZ^{32}_3&=0=\CZ^{42}_4=\CZ^{23}_2=\CZ^{24}_4~, \\ 
\CZ^{jk}_i&=0~~ (i\neq j\neq k)~. 
\label{bpsequations}\end{align}

The natural generalization of the ABJM model to a class of $\CN=3$ necklace $A_{r-1}$
quiver theories is studied in~\cite{Imamura:2008nn,Jafferis:2008qz,Herzog:2010hf}.
The dual M-theory backgrounds for these models are $AdS_4\times Y_7$,
where $Y_7$ is a seven-dimensional tri-Sasaki Einstein space which is the
base of a hyper-K\"ahler cone.

\subsection{Gaiotto-Witten matrix model\label{GWmodel}} 

A final example which we shall briefly make reference to in the
following is the Gaiotto-Witten model~\cite{Gaiotto:2008sd} which is
based on the quiver

\begin{equation}
\vspace{10pt}
\begin{xy}
\xymatrix@C=30mm{
\bullet \ar@/^1pc/[r] \ar@/_1pc/[r] & \bullet
}
\end{xy}
\vspace{10pt}
\label{quiverGW}\end{equation}
We shall not spell out all details of this reduced model, which are
completely analogous to the ABJM matrix model; indeed, we will regard
the Gaiotto-Witten matrix model as a certain reduction of the matrix
model of \S\ref{ABJMmatrix} which is schematically obtained by
removing two of the arrows from the ABJM quiver (\ref{quiverC3Z3}). In
this case there are two bifundamental matter fields and the R-symmetry
is reduced from $\sSU(4)=\sSO(6)$ to
$\sSO(4)=\sSU(2)\times\sSU(2)$. This model is also deformed by a
quartic superpotential of the chiral superfields and possesses $\CN=4$
supersymmetry; its BPS equations are identical in form to those of the ABJM
matrix model and are determined by the quantities $\CZ_i^{12}$ from
(\ref{abjmbeta}) with $i=1,2$.

\section{3-algebra models\label{3algebramodels}}

In this section we describe the 3-algebra structures underlying the Chern-Simons
quiver matrix models from \S\ref{dimredCS}, which we will use 
to analyse their vacuum structure. The various formulations of these models are related to each
 other, and it is possible to pass between them when certain
 constraints are placed on the relevant 3-algebras. For a particular
 3-algebra, we show that it is possible to pass from a certain reduced
 3-Lie algebra model to our ABJM matrix
 model.  Furthermore, in a certain scaling limit, one can reach the
 3-algebra model from the ABJM matrix model from \S\ref{ABJMmatrix}, again for a particular 3-Lie algebra.

\subsection{Reduced 3-Lie algebra model\label{reduced3Lie}} 

Let us first recall the reduced 3-Lie algebra model derived in
\cite{DeBellis:2010sy}, which is the dimensional reduction of the BLG
theory. It is a Chern-Simons-matter theory with matter fields taking
values in a metric 3-Lie algebra $\CA$ and scalars $A_\mu $ valued in
the associated Lie algebra $\frg_\CA$. The matter fields consist of
eight scalars $X^I $, $I=1,\ldots,8$, which transform under the
R-symmetry group ${\sf SO}(8)$, together with their superpartners
which can be combined into a Majorana spinor $\Psi$ of $\sSO(1,10)$
satisfying $\Gamma_{012}\Psi=-\Psi$; throughout we denote
$\Gamma_{M_1\cdots M_k}:= \frac1{k!}\, \Gamma_{[M_1}\cdots
\Gamma_{M_k]}$ where $\Gamma^I$, together with the gamma-matrices
$\Gamma^\mu$, $\mu=0,1,2$, form the generators of the Clifford algebra
$C\ell(\FR^{1,10})$. The Chern-Simons term is constructed using the
alternative invariant form $\Tr_{\frg_\CA}$ available on $\frg_\CA$
from \eqref{TrgA}. The
action reads as
\begin{equation}\label{3liealgebramodel}
\begin{aligned}
 \mathcal{S}_{\rm BLG}=&\ \tfrac{1}{6}\, \epsilon^{\mu\nu\lambda}\, \Tr_{\frg_\CA}\big( A_\mu
 \, [A_\nu,A_\lambda]\big)-\tfrac{1}{2}\, (A_\mu  X^I
 ,A^\mu  X^I)+\tfrac{\di}{2}\, (\bar{\Psi}, \Gamma^\mu  A_\mu  \Psi)\\
&\ +\tfrac{\di}{4}\, (\bar{\Psi}, \Gamma_{IJ}
[X^I,X^J,\Psi])-\tfrac{1}{12}\, ([X^I,X^J,X^K],[X^I,X^J,X^K])~.
\end{aligned}
\end{equation}
This action is invariant under the $\CN=8$ supersymmetry transformations
\begin{align} \nonumber
\delta X^I&=\di\, \bar{\eps}\, \Gamma^I\, \Psi~, \\[4pt] \nonumber
\delta\Psi&=A_\mu X^I\, \Gamma^\mu\,
\Gamma_I\, \eps-\tfrac{1}{6}\, [X^I,X^J,X^K]\, \Gamma_{IJK}\, \eps~, \\[4pt] 
\delta A_\mu &=\di\, \bar{\eps}\, \Gamma_\mu\, \Gamma_I\, [X^I ,
  \Psi,-] ~.
\label{3algsusy}\end{align}
It is also invariant under the gauge transformations generated by
$\Lambda\in\mathfrak{g}_{\mathcal{A}}$ given as
\begin{align}
A_\mu \ \longmapsto \ -[A_\mu ,\Lambda]~, \qquad X^I \ \longmapsto \ \Lambda X^I~, \qquad \Psi \ \longmapsto \
\Lambda\Psi~.
\end{align}

The vacuum moduli space $\frM^{\rm BLG}_\CA$ of the 3-Lie algebra model is defined by
setting $A_\mu=0=\Psi$ and
\beq
\big[X^I,X^J,X^K\big]=0
\eeq
in order to satisfy the BPS equations implied by (\ref{3algsusy}). For
the 3-Lie algebra $\CA=A_4$, the moduli
space is given by~\cite{Distler:2008mk}
\beq
\frM^{\rm BLG}_{A_4}=\big(\FR^8/\RZ_2\big)\times\big(\FR^8/\RZ_2\big) \ .
\eeq

\subsection{Hermitian 3-algebra model\label{herm3Lie}}

Alternative 3-algebra 
models can be written down if one relaxes the requirements of maximal
supersymmetry and of total antisymmetry of the 3-bracket. 
We first break the $\sSO(8)$ R-symmetry group of
the maximally supersymmetric theory to ${\sf SU}(4)\times {\sf
  U}(1)$. The supercharges transform under $ {\sf SU}(4)\cong \sSO(6)$, whilst the
${\sf U}(1)$ factor provides an additional global symmetry. Introduce four complex 3-algebra valued
scalar fields $Z^i$, $i=1,2,3,4$. Denote the
corresponding four fermions by $\psi^i$; they are two-component Dirac
spinors of ${\sf SO}(1,2)$.  We select a real set of gamma-matrices
$\gamma_\mu$, with $\gamma_{012}=\unit$. The Majorana condition is
$\bar{\eps}=\eps^\top\, \gamma_0$. For a generic hermitian 3-algebra $\CA$,
the analog of our 3-Lie algebra model
\eqref{3liealgebramodel} is given by
\begin{align} \nonumber
\mathcal{S}^\FC_{\rm BLG} =& \ \tfrac{1}{6}\, \epsilon^{\mu\nu\lambda}\,
\Tr_{\frg_\CA}\big( A_\mu \, [A_\nu,A_\lambda]\big) -(A_\mu
Z^\dag_i,A^\mu Z^i)+\di\, (\bar{\psi}_i,\gamma^\mu \, A_\mu \psi^i) -
\CV(Z) \\ \nonumber
&\ -(\bar{\psi}{}^i,[\psi_i,Z^j;Z_j]) +2\, \di\,
(\bar{\psi}{}^i,[\psi_j,Z^j;Z_i])+\tfrac{\di}{2}\,
\epsilon_{ijkl}\, (\bar{\psi}{}^i,[Z^k,Z^l;\psi^j]) \\ &\ -\tfrac{\di}{2}\,
\epsilon_{ijkl}\, (Z^l,[\bar{\psi}{}^i,\psi^j; {Z}^k])~,
\label{complex3algebramodel}\end{align}
where the sextic potential is given by
\begin{align}\label{complexpotential}
\CV(Z)&=\tfrac{2}{3}\,\big(\Upsilon^{jk}_i(Z)\,,\,\Upsilon^{jk}_i(Z)^\dag
\big)
\end{align}
with
\begin{align}
\Upsilon^{jk}_i(Z)&=[Z^j,Z^k;Z_i]-\tfrac{1}{2}\, \delta_i^j\,
[Z^l,Z^k;Z_l]+\tfrac{1}{2}\, \delta^{k}_i\, [Z^l,Z^j;Z_l ]~.
\end{align}
The supersymmetry transformations of this model read
\begin{align} \nonumber
\delta Z^i&=\di\, \bar{\eps}{}^{ij}\, \psi_j~, \\[4pt] \nonumber
\delta\psi^i&=-\gamma^\mu  \, A_\mu  Z_j \, \varepsilon^{ij}+[Z_j,Z^k;Z_k
]\, \varepsilon^{ij}+[Z^k,Z^l;Z^i
]\, \varepsilon_{kl}~, \\[4pt] 
\delta A_\mu &=-\di\, [-,Z_i;\psi_j] \, \gamma_\mu\, \varepsilon^{ij}+\di\,
{\bar\varepsilon}{}^{ij} \,
\gamma_\mu \, [-,\psi_j ; Z_i] ~.
\end{align}

We will now parallel the construction of~\cite{Bagger:2008se} to
demonstrate that our reduced model, for a particular choice of
hermitian 3-algebra $\CA$ and gauge group, yields the $\CN=6$ ABJM
matrix model. Let $\mathcal{A}=\sHom(V_L,V_R)$
with 3-bracket \eqref{hermitianbracket} and inner product
\eqref{hermitianinnerproduct}. The gauge group is the product ${\sf
  U}(N_L)\times {\sf U}(N_R)$, corresponding to the associated Lie algebra
$\frg_\CA= \mathfrak{u}(N_L)\oplus\mathfrak{u}(N_R)$.
With these choices, the action \eqref{complex3algebramodel} becomes
\begin{align} \nonumber
\mathcal{S}_{\rm BLG}^\FC=&\ \Tr_V\Big(A^L_\mu\, Z^\dag_i\, Z^i\,
A^L_\mu +A^R_\mu\, Z^i\, Z^\dag_i\, A^R_\mu -2A^L_\mu\, Z^\dag_i\,
A^R_\mu \, Z^i-\di\, \bar{\psi}{}^i\, \gamma^\mu\, A^R_\mu \, \psi_i
+\di\, \bar{\psi}{}^i\, \gamma^\mu \, \psi_i\, A^L_\mu \\
\nonumber &\qquad \qquad +\tfrac{1}{6}\, \epsilon^{\mu\nu\lambda}\,
\big(A^L_\mu \, \big[A^L_\nu,A^L_\lambda\big]-A^R_\mu \,
\big[A^R_\nu,A^R_\lambda\big] \big)-\CV(Z) \\ \nonumber & \qquad \qquad
-\di\, \lambda\, \big( \bar\psi{}^{i}\,
\psi_{i} \, Z_j^\dag \, Z^j+ \bar\psi{}^{i} \, Z^j\, Z_j^\dag \,
\psi_{i} -2\bar\psi{}^{i}\, \psi_{j}\, Z^\dag_i \, Z^j+ \bar\psi{}^{i} \, Z^j\,
Z_i^\dag \, \psi_{j}\big) \\ & \qquad \qquad +\di\, \lambda\, \big( \epsilon_{ijkl}\, \bar\psi{}^{i}\,
Z^k\, \bar\psi{}^{j}\,  Z^l - \epsilon^{ijkl}\, Z_l^\dag \,
{\psi}_i\, Z_k^\dag \, \psi_j\big) \Big)~.
\end{align}
Our choice of 3-bracket is antisymmetric in the first two entries. This lets us rewrite the potential \eqref{complexpotential} as
\begin{align}
\CV(Z)=\Tr_V\big(-\tfrac{2}{3}\, \big[Z^i,Z^j;Z^k\big]\,
\big[Z^\dag_i,Z^\dag_j;Z^\dag_k\big]+\tfrac{1}{2}\, \big[Z^k,Z^i;Z^i \big]\,
\big[Z^\dag_k,Z^\dag_j;Z^\dag_j\big] \big)~.
\end{align}
The global minima of $\CV(Z)$ are described by the equations
\begin{align}
\CZ^{jk}_i:=\big[Z^j,Z^k;Z_i \big]=0~,
\label{VZglobalmin}\end{align}
which are just the BPS equations \eqref{bpsequations} with $A_\mu=0$. These equations coincide with the extrema of
the superpotential \eqref{ABJMsuperpot}, and
hence define the relations of the double of the ABJM quiver (\ref{quiverC3Z3}).
We can evaluate the 3-brackets explicitly, and then the potential
assumes the manifestly $\sSU(4)$-invariant form
\begin{align}\nonumber
\CV(Z)=&\ - \tfrac{2\lambda^2}{3}\, \Tr_V \big(2Z^k\, Z^\dag_j\, Z^i\, Z^\dag_k\,
Z^j\, Z^\dag_i+2Z^k\, Z^\dag_j\, Z^i\, Z^\dag_i\, Z^j\, Z^\dag_k \\ 
& \qquad \qquad \qquad +\tfrac{1}{2}\, Z^i\, Z^\dag_i\, Z^k\, Z^\dag_k\, Z^j\,
Z^\dag_j+ \tfrac{1}{2}\, Z^\dag_i\, Z^i\, Z^\dag_j\, Z^j\, Z^\dag_k\,
Z^k\big) \ .
\end{align}
For the choice of constant $\lambda=\frac1{2\kappa}$, we recover the
$\mathcal{N}=6$ ABJM matrix model \eqref{reducedabjm}.

Note that the BPS equations
(\ref{VZglobalmin}) and their conjugates imply that the collection
$Z_i^\dag\, Z^j$ of endomorphisms of $V_L$ for $i,j=1,2,3,4$ form a mutually
commuting set of $N_L\times N_L$ matrices; similary $Z^j\, Z_i^\dag$
are a mutually commuting set of $N_R\times N_R$ matrices. 
In the ABJM limit $N_L=N_R=N$, the operators $Z_i^\dag\, Z^j$ and $Z^j\, Z_i^\dag$
moreover have the same spectra, and the vacuum
moduli space $\frM^{\rm ABJM}_{N}$ is therefore given by the $N$-th symmetric
product orbifold
\beq
\frM^{\rm ABJM}_{N}= \big(\FC^4\big)^N\, \big/\, \frS_N
\label{frMNSymN}\eeq
where $\frS_N$ is the Weyl group of $\sU(N)$ acting by permuting
the components of $N$-vectors. As we will make use of this result
later, let us derive it explicitly. For this, we note that the BPS
equations in this case are solved by commuting matrices $[Z^i,Z^j]=0$,
$i,j=1,2,3,4$. Then $Z^i$ can be put simultaneously into their Jordan
normal forms, with $k$ eigenvalues $\zeta_1^i,\dots,\zeta_k^i$ of each
endomorphism $Z^i$, i.e. for each fixed $i\in\{1,2,3,4\}$, each
$\zeta_l^i$, $l=1,\dots,k$, corresponds to a Jordan block; doing so
breaks the $\sU(N)\times\sU(N)$ gauge symmetry to a diagonal $\sU(N)$ subgroup. To every
Jordan block one associates its dimension $\lambda_l$, independently
of $i\in\{1,2,3,4\}$ because $Z^i$ mutually commute. The collection
$\lambda=(\lambda_1,\dots,\lambda_k)$ of dimensions satisfies
\beq
\lambda_1\geq\lambda_2\geq\cdots\geq\lambda_k\geq0 \ , \qquad
\sum_{l=1}^k\, \lambda_l=N \ ,
\eeq
and thus defines a linear partition of the rank $N$ of length $k$. Then the
isomorphism (\ref{frMNSymN}) is generated by the map
\beq
\big(Z^1,Z^2,Z^3,Z^4\big) \ \longmapsto \ \sum_{l=1}^k\, \lambda_l\,
\vec z_l \ \in \ \big(\FC^4\big)^N\, \big/\, \frS_N \ ,
\label{ZiSNmap}\eeq
where $\big\{\vec
z_l=(\zeta_l^1,\zeta_l^2,\zeta_l^3,\zeta_l^4)\big\}_{l=1,\dots,k}$ is
a set of $k$ points in $\FC^4$ with multiplicities given by the linear
partition $\lambda$.

A completely analogous calculation shows that the moduli space of
vacua $\frM_N^{\rm GW}$ of the $\sU(N)\times \sU(N)$ Gaiotto-Witten
matrix model is the orbifold
\beq
\frM_N^{\rm GW}=\big(\FC^2\big)^N\, \big/\, \frS_N \ .
\eeq
In this case the eigenvalues of the commuting bilinear matrices $Z^j\, Z_i^\dag$,
$i,j=1,2$, can be interpreted as points in $\FR^4=\FC^2$.

For more general quiver matrix models, the moduli space of vacua
typically contains the symmetric product $X^N/\frS_N$, where $X$ is an
affine Calabi-Yau fourfold; for a large class of such theories, $X$ is
topologically a cone over a compact Sasaki-Einstein seven-manifold $Y_7$
(see e.g.~\cite{Martelli:2008si}).
 
 \subsection{Mapping to the lorentzian Lie algebra model}

Following~\cite{Honma:2008jd}, we shall now demonstrate how a particular
contraction relates the lorentzian version of the 3-Lie algebra model
\eqref{3liealgebramodel} with the ABJM matrix model \eqref{reducedabjm}.
The first step is to construct the \emph{lorentzian Lie algebra
  model}. We fix a semisimple Lie algebra $\frh$ and
expand the fields of the reduced 3-Lie algebra model in terms of the
generators of $\CA_\frh$ satisfying the 3-bracket relations \eqref{nappiwitten3liealgebra} as
\begin{align}
X^I&=X_{\rm c}^I\, \unit+X^I_0\, \tau_0 +X^I_a\, \tau_a~, \nonumber
\\[4pt] \Psi&=\Psi_{\rm c}\,\unit+\Psi_0 \, \tau_0+\Psi_a\, \tau_a~, \nonumber \\[4pt]
A^\mu &=A^\mu_{0{a}}\, D_{0 {a}}+A^\mu_{ab}\, D_{ab}~.
\end{align}
It is convenient to make the field definitions
\begin{align}
\hat{X}^I=X^I_a\, \tau_a~, \qquad \hat{\Psi}=\Psi_a\, \tau_a~, \qquad
\hat{A}{}^\mu =A^\mu_{0a}\, \tau_{a}~, \qquad B^\mu
=f_{abc}\, A^\mu _{ab}\, \tau_c~.
\end{align}
We insert these expansions into \eqref{3liealgebramodel}, and
denote the inner product \eqref{nappiwitteninnerproduct} by $\Tr_{\frh}$
here. Using \eqref{reducedbracket}, 3-brackets involving the generator
$\tau_0$ induce the Lie bracket of $\CA_\frh'$ through
\begin{align}
\big[X^I,X^J,\tau_0 \big]=\big[\hat{X}^I,\hat{X}^J\big]~.
\end{align}
 A similar reduction occurs for the brackets involving fermions. For
 the terms involving the gauge fields, we use
 (\ref{nappiwitteninnerproduct}) to infer that terms proportional to
 the central element
 $\unit$ decouple from the gauge interactions, and in fact completely
 from the action. In this way we find the lorentzian Lie algebra model
\begin{align}\nonumber
\mathcal{S}_{\frh}=&\ \Tr_{\frh}\Big(\big(\tfrac{1}{2}\, [\hat{A}^\mu
,\hat{X}^I]+B^\mu \,
X^I_0\big)^2+\tfrac{1}{4}\,(X^K_0)^2\,
[\hat{X}^I,\hat{X}^{J}]^2-\tfrac{1}{2}\, \big(X^I_0\,
[\hat{X}^I ,\hat{X}^J]\big)^2 \\ \nonumber
& \qquad \qquad  +\tfrac{\di}{2}\,\bar{\hat{\Psi}}\,\Gamma^\mu \, [\hat{A}_\mu
,\hat{\Psi}] -\di\, \bar{\Psi}_0\, \Gamma^\mu \, B_\mu \,
\hat{\Psi}-\tfrac{1}{2}\, \bar{\Psi}_0\, \hat X^I\,
[\hat X^J,\Gamma_{IJ} \hat{\Psi}] +\tfrac{1}{2}\, \bar{\hat{\Psi}}\,
X^I_0\, [\hat X^J,\Gamma_{IJ}\hat{\Psi}] \\ & \qquad \qquad +\tfrac{1}{2}\,
\epsilon^{\mu\nu\lambda}\, [\hat A_\mu ,\hat A_\nu]\, B_\lambda \Big)~.
\label{lorentzianmodel}\end{align}
It is invariant under the supersymmetry transformations
\begin{align} \nonumber
\delta\hat{\Psi}&=\big([\hat{A}^\mu ,\hat{X}^I]+B^\mu \, X^I_0\big)\, \Gamma_\mu\,
\Gamma_I\, \eps-\tfrac{1}{2}\, X^K_0\, [\hat{X}^I,\hat{X}^J]\,
\Gamma_{IJK} \,
\eps~, \qquad \delta\Psi_0=0~,\\[4pt] \nonumber
\delta\hat{X}^I&=\di\, \Gamma^I\, \bar{\eps}\, \hat{\Psi}~, \qquad
\delta X^I_0=\di\, \Gamma^I\, \bar{\eps}\, \Psi_0~, \\[4pt]
\delta B^\mu &=\di\, \bar{\eps}\, \Gamma^\mu\, \Gamma_I\, [\hat{X}^I,
\hat{\Psi}]~, \qquad \delta\hat{A}^\mu =\di\, \bar{\eps}\, \Gamma^\mu\,
\Gamma_I\, \hat{X}^I\, \Psi_0+\di\, \bar{\eps}\, \Gamma^\mu\,
\Gamma_I\, X^I_0\, \hat{\Psi}~.
\end{align}
In the following we show how this model is related to the ABJM
matrix model \eqref{reducedabjm}: For a particular choice of gauge symmetry
breaking and scaling limit, we show that one can recover the
lorentzian Lie algebra model (\ref{lorentzianmodel}) from
(\ref{reducedabjm}). As we will make use of similar reductions
throughout this paper, we describe it here in detail.

For this, we consider the ABJM limit $N_L=N_R=N$. To take the scaling limit, we first make
the gauge field redefinitions
\begin{align}\label{axialgauge}
A^L_\mu =A_\mu +\di\, B_\mu ~, \qquad A^R_\mu =A_\mu -\di\, B_\mu
\end{align}
which breaks the gauge symmetry to a diagonal $\sU(N)$ subgroup of
$G=\sU(N)\times \sU(N)$. With this replacement, the Chern-Simons term
from the first line of (\ref{reducedabjm}) reads
\begin{align}
\mathcal{S}_{\rm g}=\kappa\, \epsilon^{\mu\nu\lambda}\,
\Tr_V\big(B_\mu \, [A_\nu,A_\lambda]-\tfrac{1}{3}\, B_\mu\, B_\nu\,
B_\lambda\big) \ .
\label{CSmatrix}\end{align}
We write the real and imaginary parts of the scalars and fermions as
\begin{align}
Z^i=X^i+\di \, X^{i+4}~, \qquad \psi^i=\chi^{i}+\di\, \chi^{i+4}
\label{Zpsirealim}\end{align}
for $i=1,2,3,4$. 
We decompose the scalars and fermions further into trace and traceless
components as
\begin{align}\nonumber
Z^i&=X_0^i\, {\tau_0}+\di \, X^{i+4}_0\, {\tau_0}+X_a^i\,
\tau_a+\di\, X_a^{i+4}\, \tau_a~, \\[4pt]
\psi^i&=\psi_0^i\, {\tau_0}+\di\, \psi_0^{i+4}\, \tau_0 +\psi_a^i\, \tau_a+\di\, \psi_a^{i+4}\, \tau_a~.
\label{Zpsitrace}\end{align}
In this decomposition we have identified ${\tau_0}$ with the generator
of $\mathfrak{u}(1)$, and $\tau_a$, $a=1,\dots,d=N^2-1$, are the generators of $\mathfrak{su}(N)$.
We scale the fields as
\begin{align}\label{lorentzianscaling}
B_\mu \ \longrightarrow \ g\, B_\mu ~, \qquad X^i_0 \ \longrightarrow \
\tfrac{1}{g}\, X^i_0~, \qquad \psi^i_0 \ \longrightarrow
\ \tfrac{1}{g}\, \psi^i_0
\end{align}
with all other fields unchanged, and the coupling constant as
$\kappa\to\frac1g\, \kappa$. Taking the limit $g\to0$ we find that the
Chern-Simons term (\ref{CSmatrix}) reduces to
\begin{align}
\mathcal{S}_{\rm g}=\kappa\, \epsilon^{\mu\nu\lambda}\,
\Tr_V\big(B_\mu\, [A_\nu,A_\lambda] \big)~,
\label{CSreduced}\end{align}
while the second line of \eqref{reducedabjm} becomes
\begin{align}
\mathcal{S}_{\rm k}=-\Tr_V\Big( \big([A_\mu ,X^I]+2B_\mu\,
X^I_0\big)^2+\di\, \bar{\psi}\, \gamma^\mu\, [A_\mu
,{\psi}]-2\bar{\psi}\, \gamma^\mu \, B_\mu\,
\psi_0-2\bar{\psi}_0\, \gamma^\mu \, B_\mu \, \psi\Big)~.
\end{align}
In this reduction we have combined the indices $i$ and $i+4$ for
$i=1,2,3,4$ into an index $I=1,\dots,8$, and the components of the spinors into a single Majorana fermion
\begin{align}\label{Majoranafermion}
\psi=\big(\chi^1,\dots,\chi^8
\big)^\top \ .
\end{align}
Now we consider the bosonic sextic potential. In the scaling limit,
the surviving terms from the potential contain four trace components
and eight real traceless components. Using $\sSU(4)$ R-symmetry we arrange them as
\begin{align}
Z^i=\delta^{i,1}\, \big(X^i_0+\di \, X^{i+4}_0\big)\,
{\tau_0}+\big(X_a^i +\di\, X_a^{i+4}\big) \, \tau_a~.
\label{Ziarrange}\end{align}
If we combine the trace components as
\begin{align}
X^I_0=\big(X^1_0,0,0,0,X^5_0,0,0,0\big)~,
\end{align}
then the reduced bosonic potential reads
\begin{align}
\CV_{\rm b}(X) =-\tfrac{1}{2\kappa^2}\, \Tr_V\Big(\tfrac{1}{4}\,
\big(X^K_0\big)^2\, \big[X^I,X^J\big]^2-\tfrac{1}{2}\,
\big(X^I_0\, [X^I,X^J]\big)^2 \Big)~.
\end{align}
We finally consider the quartic Yukawa potential. In this scaling
limit, the surviving term of this potential has contributions from two
bosonic trace components and two traceless bosonic components. We
arrange them as in (\ref{Ziarrange}) and the spinor components into a
Majorana fermion as in \eqref{Majoranafermion}. The resulting potential reads
\begin{align}
\CV_{\rm f}(X,\psi)=-\tfrac{1}{\kappa}\, \Tr_V\big(\bar{\psi}\,
X^I_0\, [X^J,\gamma_{IJ}\, \psi]\big)
\end{align}
for suitable antisymmetrized products of $8
\times 8$ gamma-matrices $\gamma_{IJ}$ (see
e.g.~\cite[App.~A]{Honma:2008jd}).

The fully contracted theory thus reads
\begin{align} \nonumber
\mathcal{S}_{\rm red} =&\ -\Tr_V\Big(\big([A_\mu ,X^I]+2B_\mu\,
X^I_0\big)^2+\di\, \bar{\psi}\, \gamma^\mu \, [A_\mu
,{\psi}]-2\bar{\psi}\, \gamma^\mu \, B_\mu \,
\psi_0-2\bar{\psi}_0\, \gamma^\mu\, B_\mu \,\psi \\ \nonumber
& \qquad \qquad
-\tfrac{1}{2\kappa^2}\,\big(\tfrac{1}{4}\, (X^K_0)^2\,
[X^I,X^J]^2-\tfrac{1}{2}\, X^I_0\,
[X^I,X^J]^2\big)-\tfrac{1}{\kappa}\, \bar{\psi}\, X^I_0\,
[X^J,\gamma_{IJ}\, \psi] \\ 
& \qquad \qquad +\kappa \, \epsilon^{\mu\nu\lambda}\, [A_\mu
,A_\nu]\, B_\lambda\Big)~.
\end{align}
This is just the original lorentzian Lie algebra model
\eqref{lorentzianmodel} with $\frh=\mathfrak{su}(N)$ and inner
product~\eqref{hermitianinnerproduct}.

The connection between the generic Gaiotto-Witten matrix models and
reduced 3-algebra models is less clear. For the reduced 3-Lie algebra
model of \S\ref{reduced3Lie} based on $\CA=A_4$, one can reduce the
$\sSO(8)$ R-symmetry group to $\sSO(4)$ by keeping only half of the
supersymmetries; then the matter content splits into a hypermultiplet
plus a twisted hypermultiplet, and the reduced model can be regarded as the
$\sSU(2)\times \sSU(2)$ Gaiotto-Witten model with an additional
twisted hypermultiplet~\cite{Hosomichi:2008jd}.

\section{Dual IKKT matrix models\label{IKKTmodels}}

A main driving force in our subsequent analysis will be certain
connections to the IKKT matrix model~\cite{Ishibashi:1996xs}, whose vacuum moduli space is
well understood and for which detailed localization techniques are
available~\cite{Moore:1998et}. In this section we exploit the fact
that it is also possible to reach gauge theories on D2-branes from
M2-brane theories; this occurs in the limit away from the orbifold
point of $\FR^8/\RZ_k$ where the orbifold geometry is $S^1\times
\FR^7$. A field theory realization of this reduction is provided by
the  Higgs mechanism of~\cite{Mukhi:2008ux}, which relates the
maximally supersymmetric BLG Chern-Simons-matter theory to $\mathcal{N}=8$ supersymmetric Yang-Mills theory in three
 dimensions: One gives a vacuum expectation value $v$ to
 a single scalar field and takes the limit $\kappa ,v\to\infty$ with
 $g^2:=\frac{v^2}{2\kappa}$ fixed; this reduction was exploited in~\cite{DeBellis:2010sy}
 to relate the reduced 3-algebra model (\ref{3liealgebramodel}) and
 its classical solutions to those of the IKKT model in ten
 dimensions. In~\cite{Sato:2013iqa} it was shown how a certain variant
 of the lorentzian Lie algebra model (\ref{lorentzianmodel}) reduces
 to the ten-dimensional IKKT matrix model in a suitable BPS background.
It was shown in \cite{Pang:2008hw} that the Higgs
 mechanism proposed by \cite{Mukhi:2008ux} also reduces the ABJM
 theory to $\mathcal{N}=8$ supersymmetric Yang-Mills theory in three
 dimensions. By considering dimensional reductions of generic
 $\CN=2$ Chern-Simons-matter theories, we can reduce via the Mukhi-Papageorgakis map to a variety of Yang-Mills matrix models. We demonstrate this explicitly for a
 particular reduction of the $A_1$ quiver matrix model to the $\CN=1$ IKKT model in
 four dimensions. We then apply this map to our ABJM matrix
 model, and arrive at the $\mathcal{N}=8$ ten-dimensional IKKT matrix
 model. In this section only we shall take the ABJM limit $N_L=N_R=N$
 throughout.

\subsection{Mapping of the $A_1$ quiver matrix model}

We begin with the simplest example for which the reduction is
relatively straightforward to construct. We consider the dimensionally reduced $\mathcal{N}=2$
Chern-Simons-matter theory \eqref{bifundamentalmatter}, and show that
under the Mukhi-Papageorgakis map it reduces to the
four-dimensional IKKT matrix model with $\mathcal{N}=1$ supersymmetry
and gauge group ${\sf SU}(N)$. We work with the Clifford algebra
$C\ell(\FR^{1,2})$, and use Dirac spinors. Our gamma-matrices are
the Pauli spin matrices, and the Majorana conditions read
\begin{align}
\bar{\eps}\, \lambda=\bar\lambda\, {\eps}~, \qquad \bar{\eps}\,
\gamma^\mu\, \lambda=-\bar\lambda\, \gamma^\mu\, {\eps}~.
\label{A1Majcond}\end{align}
As previously, we break the gauge symmetry to a $\sU(N)$ subgroup by
making the field replacements \eqref{axialgauge}. We also restrict the
matter field $Z$ to be hermitian. We decompose $Z$ into components
$Z'\in \mathfrak{su}(N)$ and $Z_0\in \mathfrak{u}(1)$, and expand it
around a classical value proportional to the identity with a coupling
constant $g$ as
\begin{align}
Z=g\, \unit+Z_0+Z^\prime~.
\end{align}
Using global $\sU(1)$ symmetry, we may take $g\in\FR$. For the gaugino and auxiliary fields, we take a diagonal limit in which
\begin{align}
\lambda^L=-\lambda^R=:\lambda~, \qquad D^L=D^R=: D~, \qquad
\sigma^L=\sigma^R=:\sigma ~,
\label{A1gauginolimit}\end{align}
and further couple the gauge and matter sectors of the model together
by the requirements
\begin{align}
\lambda=-g\, \psi~, \qquad \sigma=g\, Z \ , \qquad D=-g\, F~.
\end{align}

With these gauge field replacements, and diagonal limits of the
gauginos and auxiliary fields, we find that the pure Chern-Simons
component from the first line of the action
\eqref{bifundamentalmatter} reduces to (\ref{CSmatrix}). For the
remaining matter contributions in \eqref{bifundamentalmatter}, by inserting the
field identifications above and expanding around the vacuum value we obtain
\begin{align}
\mathcal{S}_{\rm m}=&\Tr_V\big(-[A_\mu ,Z^\prime\, ]^2-4g^2\,
B_\mu\,B^\mu \nonumber \\ & \qquad \qquad +
\di\, \bar{\psi}\,\gamma^\mu\, [A_\mu ,\psi]-\bar{\psi}\, \gamma^\mu\, \{B_\mu
,\psi\}+ \di\,g\, \bar{\psi}\, [Z^\prime, \psi]+F^2\big)~.  
\label{A1matterred}\end{align}
We now scale the fields appropriately and take the strong coupling
limit $g\to\infty$. We can integrate out the auxiliary field $B_\mu $
using its equation of motion
\begin{align}
B_\mu =\tfrac{\kappa}{g^2}\, \epsilon_{\mu\nu\lambda}\, [A^\nu ,A^\lambda]~.
\label{BmuA1}\end{align}
In deriving this equation we have ignored cubic and higher order
interactions involving $B_\mu$ that become suppressed in the strong
coupling limit. Inserting (\ref{BmuA1}) into the pure Chern-Simons
action (\ref{CSmatrix}), we find
\begin{align}
\mathcal{S}_{\rm g}=-\tfrac{4\kappa^2}{g^2}\, \Tr_V\big([A_\mu
,A_\nu]^2\big) ~.
\label{SgA1}\end{align}
We scale the matter field $Z$ by the factor $\tfrac{1}{g}$, and
similarly for the matter fermion (and its adjoint) and the auxilliary
field $F$. Replacing $B_\mu $ by its equation of motion (\ref{BmuA1}), we find that
the matter action (\ref{A1matterred}) reduces in the strong coupling
limit to
\begin{align}
\mathcal{S}_{\rm m}=\Tr_V\big(-\tfrac{1}{g^2}\, [A_\mu
,Z^\prime\,]^2-\tfrac{4\kappa^2}{g^2}\, [A_\mu ,A_\nu]^2+
\tfrac{\di}{g^2}\, \bar{\psi}\, \gamma^\mu \, [A_\mu
,\psi]+\tfrac{\di}{g^2}\, \bar{\psi}\,
[Z^\prime,\psi]+\tfrac{1}{g^2}\, F^2\big) ~. 
\label{SmA1}\end{align}

We combine the scalar and gauge fields into a single field 
\begin{align}
X^I=(X^\mu,X^3)=(A^\mu,Z'\,)
\label{A1XI}\end{align}
where $I=0,1,2,3$. Then with $\kappa=\tfrac{1}{4}$, the sum of
(\ref{SgA1}) with the first two terms of (\ref{SmA1}) can be written
as $-\tfrac{1}{2g^2}\, \Tr_V\big([X^I,X^J]^2\big)$, which is the bosonic potential of the IKKT model.
For the last three terms of (\ref{SmA1}), we define a four-dimensional
Majorana spinor of the Clifford algebra $C\ell(\FR^{1,3})$ by
\begin{align}
\Psi=\big(\psi^{1},\psi^{2} \big)^\top~,
\end{align}
where each real component $\psi^1$, $\psi^2$ of the Dirac spinor
$\psi$ is a two-component Majorana spinor. We then construct a set of
four-dimensional gamma-matrices from our three-dimensional Pauli spin
matrices as
\begin{align}
\Gamma^\mu =\di\,\begin{pmatrix} 0 & \gamma^\mu  \\-\gamma^\mu  &
  0\end{pmatrix} \ , \qquad \Gamma^3=-\di\, \begin{pmatrix} 0 & \unit
  \\\unit & 0 \end{pmatrix} \ .
\label{A1Gamma4D}\end{align}
For the chirality and charge conjugation matrices,  we take
\begin{align}
\Gamma^5=\begin{pmatrix}\unit & 0 \\0 & -\unit\end{pmatrix}~, \qquad 
C= \begin{pmatrix} -\di\, \gamma^2 & 0 \\0 & \di\,
  \gamma^2\end{pmatrix} ~.
\label{A1Gamma5C}\end{align}
We can then combine the last three terms of (\ref{SmA1}) as 
$\frac1{g^2}\, \Tr_V\big(-\bar{\Psi}\, \Gamma^I\,
[X_I,\Psi]+F^2\big)$, which is the fermionic term of the IKKT model together with an auxiliary field.

Altogether the $A_1$ quiver matrix model action is reduced under the
Mukhi-Papageorgakis map to the action of the four-dimensional IKKT model
\begin{align}
\mathcal{S}_{\rm IKKT}=\tfrac{1}{g^2}\, \Tr_V\big(-\tfrac{1}{2}\,
[X^I,X^J]^2-\bar{\Psi}\, \Gamma^I\, [X_I ,\Psi]+F^2\big)~.
\label{4DIKKT}\end{align}

\subsection{Supersymmetry reduction}

We will now show explicitly how the supersymmetry
transformations of the $A_1$ quiver matrix model map to those of the IKKT
model under the
Mukhi-Papageorgakis map. The original matrix model has
$\mathcal{N}=2$ supersymmetry, while the IKKT model in four dimensions
has $\mathcal{N}=1$ supersymmetry. Hence the scaling limit must reduce
the supersymmetry; we do this by identifying the infinitesimal
supersymmetry generators in (\ref{A1susy}) so that $\eps={\eta}$ are no longer independent.
We demonstrate the reduction on each field transformation of
(\ref{A1susy}) individually. 

For the transformations of the gauge
fields $A_\mu$ in (\ref{A1susy}), we make the gauge field
identifications, identify the supersymmetry generators with each
other, and scale the spinor.  In four dimensions we write the fermions
as four-component Majorana spinors obeying (\ref{A1Majcond}), and along with
the four-dimensional gamma-matrices (\ref{A1Gamma4D}) we can write
\begin{align}
\delta A_\mu =\bar{\eps}\, \Gamma_\mu\, \lambda~.
\label{A1deltaAmu}\end{align}
Following a similar process for the supersymmetry transformations of
the matter field $Z$, the requirement (\ref{A1gauginolimit}) reveals
the Majorana spinor condition (\ref{A1Majcond}). After expanding $Z$
around its classical value and scaling, we can combine its
supersymmetry transformation with (\ref{A1deltaAmu}) to get
\begin{align}
\delta X^I=\bar{\eps}\, \Gamma^I\, \Psi~.
\end{align}

For the supersymmetry variation of the auxiliary field $D$ in
(\ref{A1susy}), we identify the supersymmetry generators with each
other, scale the fields and expand around the classical value, so that
the resulting supersymmetry transformation reads as
\begin{align}
\delta D=\tfrac{\di}{2}\, \big(\bar{\eps}\, \gamma^\mu \, [A_\mu
,\lambda]+[A_\mu ,\bar{\lambda}]\, \gamma^\mu \,
\eps\big)+\tfrac{\di}{2}\, \big(\bar{\eps}\,
[\lambda,Z^\prime\,]+[\bar{\lambda},Z^\prime\, ]\, \eps\big)~.
\end{align}
Identifying the four-dimensional gamma-matrices (\ref{A1Gamma4D}) and
(\ref{A1Gamma5C}), applying the Majorana spinor identity
(\ref{A1Majcond}), and combining $Z^\prime$ with $A_\mu $ as in
(\ref{A1XI}) results in the transformation
\begin{align}
\delta D=\di\, \bar{\eps}\, \Gamma^5\, \Gamma^I\, [X_I,\Psi]~.
\label{Dsusyred}\end{align}
A similar modification occurs for the supersymmetry transformation of
the auxiliary field $F$. We take an axial combination
(\ref{axialgauge}) of the gauge fields in (\ref{A1susy}) which reduces
the interaction of the fermion and the gauge field to a commutator, at
the cost of introducing the field $B_\mu $, so that after making the field
replacements (\ref{A1gauginolimit}) we arrive at
\begin{align}
\delta F=&\ \bar{\eps}\,{}^\ast\big(\gamma^\mu \, [A_\mu
,\psi]+ \di\,\gamma^\mu\, \{B_\mu,\psi\}
+[\sigma,\psi] +\di\,\{\lambda,Z \}\big) ~.
\label{Fsusyred}\end{align}
Expanding around the vacuum, and
taking the appropriate scaling limit, the $B_\mu $ contribution
decouples. After combining the gauge and matter fields, and
rewriting the spinor and gamma-matrices, the reduction
(\ref{Fsusyred}) coincides with (\ref{Dsusyred}). 

Finally, we consider the spinor supersymmetry transformations. For the
gaugino variation $\delta\lambda$ in (\ref{A1susy}), we make the usual field
identifications and scalings, and combine the terms involving $A_\mu $
and $Z$ to get
\begin{align}
\delta \Psi=&\ -\di\,\Gamma_{IJ}\, [X^I,X^J]\, \eps-F\, \eps~.
\label{A1lambdasusyred}\end{align}
For the matter fermions in (\ref{A1susy}), we take the axial limit of the
gauge fields and make the field replacements to get
\begin{align}
\delta\psi=&\ \di\, \gamma^\mu \, \eps\, \big([A_\mu ,Z]+\di\, \{B_\mu
,Z\}\big )+F\, \eps^* ~.
\end{align}
Inserting the equation of
motion \eqref{BmuA1} for $B_\mu $ and taking the scaling limit we find
\begin{align}
\delta\psi=&\ \di\, \gamma^\mu \, \eps\, \big([A_\mu
,Z]+2\,\di\,\kappa\, \epsilon_{\mu\nu\lambda}\, [A^\nu,A^\lambda]\big)+F \, \eps~.
\label{A1psisusyred}\end{align}
By setting $\kappa=\frac14$ and using the Pauli spin matrix identity
\begin{align}
\tfrac{\di}{2} \, \epsilon^{\mu\nu\lambda}\, \gamma_\lambda
=\gamma^{\mu\nu} \ ,
\end{align}
we find that (\ref{A1psisusyred}) coincides with (\ref{A1lambdasusyred}).

\subsection{Mapping of the ABJM matrix model}

Let us now extend the
Mukhi-Papageorgakis map to reduce the ABJM model
(\ref{reducedabjm}). We break the product gauge group $G=\sU(N)\times \sU(N)$ to a
diagonal ${\sf U}(N)$ subgroup by taking an axial combination of the
gauge fields \eqref{axialgauge}. We write the real and imaginary parts
of the scalars and the spinors as in (\ref{Zpsirealim}). We further decompose the
fields into the generators of the $\sU(N)$ gauge group exactly as in
(\ref{Zpsitrace}). We expand the scalar fields around a fixed vacuum
configuration proportional to a coupling constant $g$. Using the
$\sSU(4)$ invariance of the matrix model, we can select the
scalar field $Z^4$ to expand around so that
\begin{align}
Z^i=\di\, g\, \delta^{i,4}\, \unit+X^i_0\, \tau_0+X^i_a\, \tau_a+\di\,
X^{i+4}_0\, \tau_0+\di\, X^{i+4}_a\, \tau_a~.
\end{align}

We first investigate the effect of the scaling limit on
the Chern-Simons matrix action from the first line of (\ref{reducedabjm}). The various terms of the action
separate into ${\sf U}(1)$ and ${\sf SU}(N)$ components, and in the
strong coupling limit $g\to\infty$ the ${\sf U}(1)$ terms decouple so
we will ignore them from now on. When we make the gauge field
replacement \eqref{axialgauge}, the Chern-Simons term reads as in
(\ref{CSmatrix}), which in the scaling limit will reduce to (\ref{CSreduced}).

The contributing terms to the reduction of the second line of
(\ref{reducedabjm}) give
\begin{align}\nonumber
\mathcal{S}_{\rm k}=&\ \Tr_V\Big(-\sum_{i=1}^4\,\big( [A_\mu , X^i]^2+[A_\mu
,X^{i+4}]^2\big)-4g\, [A_\mu ,X^8]\, B^\mu -4g^2\, B_\mu \, B^\mu  \\ 
& \qquad \qquad -\di\,\sum_{i=1}^4\, \big(\bar{\chi}_i\, \gamma^\mu\, [A_\mu ,\chi^i+\di \,
\chi^{i+4}]+\di\, \bar{\chi}_{i+4}\, \gamma^\mu \, [A_\mu ,\chi^i+\di \,
\chi^{i+4}] \big)\Big) \ .
\label{kineticreduced}\end{align}
Combining (\ref{CSreduced}) and (\ref{kineticreduced}),  we can
integrate out $B_\mu $ using its equation of motion
\begin{align}
B_\mu =-\tfrac{1}{g}\, [A_\mu ,X^8]+\tfrac{\kappa}{g^2}\,
\epsilon_{\mu\nu\lambda}\, [A^\nu,A^\lambda]~.
\end{align}
This causes the scalar field $X^8$ to decouple from the action. We
write the minimal spinor of $\sSO(1,2)\times\sSO(7)$ for the reduced theory as in (\ref{Majoranafermion}),
where each $\chi^{I}$ is also a two-component Majorana spinor. Then
the action (\ref{kineticreduced}) reduces to
\begin{align}
\CS_{\rm k} =& \ \Tr_V\Big(-\sum_{a=1}^3\, \big([A_\mu ,X^a]^2+ [A_\mu
,X^{a+4}]^2\big) -[A_\mu,X^4]^2 \nonumber \\ & \qquad \qquad -8\kappa^2\,
[A_\mu ,A_\nu]^2-\di\, \bar{\psi}\, \gamma^\mu\, [A_\mu ,\psi]\Big)~.
\label{kineticreduction}\end{align}

We now investigate the potential terms from the last four lines of
(\ref{reducedabjm}). The surviving terms from the bosonic potential are of the form
\begin{align}
\CV_{\rm b}(X) =& \ -\tfrac{1}{8\kappa^2}\,
\Tr_V \Big(\, \sum_{a,b=1}^3\, \big( [X^a,X^b]^2 +[X^{a+4},X^{b+4}]^2
\big) \nonumber \\ & \qquad \qquad \qquad +2\sum_{a=1}^3\, \big(
[X^a,X^4]^2+2[X^{a+4},X^4]^2 \big)\, \Big)~.
\end{align}
The fermions produce a potential that reads as
\begin{align}
\CV_{\rm f}(X,\psi)=-\tfrac{1}{\kappa}\, \Tr_V\Big(\, \sum_{a=1}^3\, \big(\bar{\psi}\, 
\gamma_a\, [X^a,\psi] +\bar{\psi}\, \gamma_{a+4} \, [X^{a+4},\psi]
\big)+\bar\psi\, \gamma_4\, [X^4,\psi]\Big) 
\end{align}
for a suitable basis of $\sSO(7)$ gamma-matrices
$\gamma^a,\gamma^4, \gamma^{a+4}$, $a=1,2,3$.

Finally, we rescale the fields as in (\ref{lorentzianscaling}), and then
the full reduced action takes the form
\begin{align}
\mathcal{S}_{\rm red}=&\ \tfrac{1}{g^2}\, \Tr_V\Big(-\sum_{a=1}^3\,
\big( [A_\mu
,X^a]^2 + [A_\mu
,X^{a+4}]^2 \big) -[A_\mu,X^4]^2-8 \kappa^2\, [A_\mu
,A_\nu]^2 \\ & \qquad \qquad -\tfrac{1}{8\kappa^2}\, \sum_{a,b=1}^3\, 
\big([X^a,X^b]^2 +[X^{a+4},X^{b+4}]^2\big) 
-\tfrac{1}{4\kappa^2}\, \sum_{a=1}^3\, \big( 
[X^a,X^4]^2+[X^{a+4},X^4]^2 \big)  \nonumber \\ 
& \qquad \qquad -\tfrac{1}{\kappa}\, \sum_{a=1}^3\, \big(\bar{\psi}\, \gamma_a\,
[X^a,\psi]+ \bar{\psi}\, \gamma_{a+4} \,
[X^{a+4},\psi] \big) -\tfrac{1}{\kappa}\,
\bar\psi\, \gamma_4\, [X^4,\psi]- \di\, \bar{\psi}\, \gamma^\mu\,
[A_\mu ,\psi] \Big)~. \nonumber
\end{align}
We can combine the bosonic fields into a single field $X^M=
(2A_\mu,X^a,X^4,X^{a+4})$ with $M=1,\dots,10$. Then this action, along with the
choice of Chern-Simons coupling constant $\kappa=\frac12$, produces
the action of the ten-dimensional IKKT matrix model. 

For later use, we note the similarity between the BPS equations of the ABJM and IKKT
matrix models. In the case of the ABJM model the BPS equations are
given by \eqref{bpsequations}, while in the case of the IKKT model the
BPS equations are determined by commuting matrices
\begin{align}
\big[X^M,X^N\big]=0~.
\label{IKKTeqs}\end{align}
However, the 3-algebra form \eqref{VZglobalmin} of the ABJM equations
does not map to the IKKT equations \eqref{IKKTeqs} under the scaling limit described
here. This is due to the removal of the gauge fields from
\eqref{bpsequations}: In the axial limit (\ref{axialgauge}) of the
gauge fields, the field $B_\mu$ causes a bosonic degree of freedom to
decouple from the action in the scaling limit in order that one may combine the gauge fields with the scalars in the
appropriate way.

A completely analogous calculation can be applied to the $\sU(N)\times
\sU(N)$ Gaitto-Witten matrix models; the Mukhi-Papageorgakis map in
this case identifies the dual theory as the $\CN=4$ six-dimensional
IKKT matrix model (see~\cite{Koh:2009um}).

\section{Cohomological 3-algebra models\label{twistedBLG}}

In what follows we shall be interested in the exact computations of the partition
functions of our Chern-Simons quiver matrix models using localization
techniques. For this, we shall need to deform our matrix models in
suitable ways in order to obtain theories with equivariant
cohomological symmetries that will enable the localization procedure
to be applied exactly. In this section we shall study cohomological
versions of our quiver matrix models that are obtained by a
topological twisting procedure, and point out various ensuing difficulties.
The possible inequivalent twists of 
Chern-Simons-matter theories in three dimensions with
$\mathcal{N}\geqslant4$ supersymmetry were classified in \cite{Koh:2009um}. In the
case of an $\mathcal{N}=8$ theory with R-symmetry group ${\sf SO}(8)$,
restricting the supercharges to the vector representation does not
generate any additional twists. However, letting the supercharges
transform in the spinor representation via the triality of the R-symmetry
group does allow for two additional twists. One of these new twists was
constructed in \cite{Lee:2008cr}; in this section we investigate
the effect of applying the
Mukhi-Papageorgakis map to this topologically twisted
theory. After dimensional reduction, the ensuing 3-algebra model
can potentially induce a cohomological
deformation of the ABJM matrix model under the mappings of
\S\ref{3algebramodels} which is dual to a novel topological twisting of the
ten-dimensional IKKT model.

\subsection{Topologically twisted BLG theory\label{BLGtwist}}

We begin by briefly reviewing the topologically twisted theory
constructed in~\cite{Lee:2008cr}. In the conventions of
\S\ref{reduced3Lie}, the BLG action in euclidean space reads
\begin{align} \nonumber
S_{\rm
  BLG}=&\ \int\, \dd^3x\ \Big( \tfrac{\di}{2}\,\epsilon^{\mu\nu\lambda}\Tr_{\frg_\CA} \big(
A_\mu\,\partial_\nu A_\lambda-\tfrac{1}{3}\, A_\mu\,
[A_\nu,A_\lambda]\big) +\tfrac{1}{2}\, (\nabla_\mu X^I,\nabla^\mu
X^I)-\tfrac{\di}{2}\, (\bar{\Psi},\Gamma^\mu\, \nabla_\mu\Psi) \\
&\qquad \qquad +\tfrac{\di}{4}\, (\bar{\Psi},\Gamma_{IJ}\,
[X^I,X^J,\Psi])+\tfrac{1}{12}\, ([X^I,X^J,X^K],[X^I,X^J,X^K]) \Big) ~.
\end{align}
This action is invariant under the 16 supersymmetries generated by
\begin{align}\nonumber
\delta X^I=&\ \di\, \bar{\eps}\, \Gamma^I\, \Psi~, \\[4pt] \nonumber
\delta\Psi=&\ \nabla_\mu X^I\, \Gamma^\mu\, \Gamma_I \, \eps
-\tfrac{1}{6}\, [X^I,X^J,X^K]\, \Gamma_{IJK}\, \eps~, \\[4pt]
\delta A_\mu=&\ \di\, \bar{\eps}\, \Gamma_\mu\, \Gamma_I\,  [X^I ,
\Psi,-] ~.
\end{align}
The main difference from the split signature case is that the
euclidean action involves only the holomorphic part of the spinor, so
that we must make the definition
\begin{align}
\bar{\Psi}:=\Psi^\top\, {\CC} \ ,
\end{align}
where ${\CC}$ is the charge conjugation matrix satisfying
\begin{align}
{\CC}\, \Gamma^M\, {\CC}^{-1}=\big(-\Gamma^M \big)^\top~, \qquad {\CC}^\top=-\CC \ ,
\end{align}
and $M$ is the 11-dimensional vector index which decomposes into
$\mu=1,2,3$ and $I=4,\dots, 11$.

Consider now the rotational symmetry breaking ${\sf Spin}(11)\rightarrow {\sf
  Spin}(3)\times {\sf Spin}(3)\times {\sf Spin}(5)$, under which the
corresponding gamma-matrices can be decomposed as
\begin{align}
\Gamma^\mu=\gamma^\mu\otimes\unit\otimes\unit\otimes\gamma^3~, \qquad
\Gamma^{\mu+3}=\unit\otimes\gamma^\mu\otimes\unit\otimes\gamma^1~,
\qquad \Gamma^{i+6}=\unit\otimes\unit\otimes\gamma^i \otimes\gamma^2
\end{align}
where $\gamma^\mu$, $\mu=1,2,3$, are Pauli spin matrices and $\gamma^i$, $i=1,\dots,5$, are $4\times4$ gamma-matrices in five euclidean dimensions. The charge conjugation matrix decomposes as
\begin{align}
\mathcal{C}=\di\, \gamma^2\otimes\di\, \gamma^2 \otimes C \otimes\unit~,
\end{align}
where $C$ is the five-dimensional charge conjugation matrix. The $\sSO(8)$ chirality matrix is
\begin{align}
\Gamma^{123}=-\di\, \Gamma^{4\cdots
  11}=\unit\otimes\unit\otimes\unit\otimes\di\, \gamma^3~.
\end{align}
This means that the spinors have four indices: two for the $\sSO(3)$
factors, one for the $\sSO(5)$ factor, and one for $\sSO(8)$
chirality. The twist is constructed by replacing an $\sSO(3)$ factor
with the diagonal subgroup of ${\sSpin}(3)\times\sSpin(3)$. Then we
can expand the twisted spinors
\begin{align} 
\Psi = (\psi,\chi^\mu)
\end{align}
into an $\sSO(3)$ scalar and vector. 
We also decompose the bosons
\begin{align}
X^I = (X^\mu,Y^i )
\end{align}
into an $\sSO(3)$ vector and five scalars.

The resulting twisted BLG action is the sum of a topological action 
\begin{align}
S_{\rm top}=& \ \int\, \dd^3x \ \Big( \tfrac{\di}{2}\, \epsilon^{\mu\nu\lambda}\,
\Tr_{\frg_\CA}\big(A^+_\mu\,\partial_\nu A^+_\lambda+\tfrac{1}{3}\,
A^+_\mu\,[A^+_\nu,A^+_\lambda]\big) \nonumber \\ &\qquad \qquad -\tfrac{1}{2} \,\epsilon^{\mu\nu\lambda}\, (
\bar{\chi}_\mu,\nabla^+_\nu\chi_\lambda-\di \, \gamma_i\,
[\chi_\nu,X_\lambda,Y^i ] ) \Big)
\label{BLGtop}\end{align}
plus a metric-dependent cohomological action
\begin{align}\label{BLGmetric}
S_{\rm m}=&\ \int\, \dd^3x\ \Big( \tfrac{1}{4}\, (\nabla^\mu
X^\nu-\nabla^\nu X^\mu,\nabla^\mu X^\nu-\nabla^\nu
X^\mu)+\tfrac{1}{2}\,
(\nabla^+_\mu Y^i,\nabla^-_\mu Y^i ) \\ \nonumber & \qquad +\tfrac{1}{2}\, (\nabla_\mu X^\mu 
+\tfrac{\di}{6}\,
\epsilon_{\mu\nu\lambda}\, [X^\mu,X^\nu,X^\lambda],\nabla_\mu X^\mu
+\tfrac{\di}{6}\, \epsilon_{\mu\nu\lambda}\,
[X^\mu,X^\nu,X^\lambda]) \\ \nonumber & \qquad +\tfrac{1}{2}\, ([Y^i ,Y^j ,Y^k
],[Y^i ,Y^j ,Y^k ]) +\tfrac{1}{2}\, ([X^\mu,Y^j ,Y^k ],[X^\mu,Y^j ,Y^k
])\\ & \qquad +(\bar{\psi},\nabla^-_\mu\chi^\mu+\di\, \gamma_i \, [Y^i
,X^\mu,\chi_\mu] +\tfrac{\di}{4}\, \gamma_{ij}\, [Y^i ,Y^j
,\psi])+\tfrac{\di}{4}\, (\bar{\chi}^\mu,\gamma_{ij}\, [Y^i ,Y^j
,\chi_\mu]) \Big)~, \nonumber
\end{align}
where the gauge fields and covariant derivatives have been complexified so that
\begin{align}
A^{\pm}_\mu :=A_\mu\mp\tfrac{\di}{2}\, \epsilon_{\mu\nu\lambda}\,
[X^\nu,X^\lambda,-]~, \qquad \nabla^{\pm}_\mu :=\nabla_\mu
\pm\tfrac{\di}{2}\, \epsilon_{\mu\nu\lambda}\, [X^\nu,X^\lambda,-]~.
\end{align}
The total action $S_{\rm top}+S_{\rm m}$ is invariant under the supersymmetry
transformations
\begin{align}
\delta X^\mu& = \bar\varepsilon\,\chi^\mu \ , \nonumber \\[4pt]
\delta Y^i&= \bar\varepsilon\, \gamma^i\, \psi \ , \nonumber \\[4pt]
\delta\psi &= -\big(\nabla_\mu X^\mu+\tfrac{\di}{6}\,
\epsilon_{\mu\nu\lambda}\, [X^\mu,X^\nu,X^\lambda]\big)\, \varepsilon
\ , \nonumber \\[4pt]
\delta \chi_\mu &= \epsilon_{\mu\nu\lambda}\, \nabla^\nu X^\lambda\, \varepsilon
+\nabla_\mu^+ Y^i\,\gamma_i\, \varepsilon +\tfrac{\di}{2}\,
[Y^i,Y^j,X_\mu]\, \gamma_{ij}\, \eps \ , \nonumber \\[4pt]
\delta A_\mu &= \di\, \bar\eps\, \big(-[X_\mu,\psi,-] +
\epsilon_{\mu\nu\lambda}\, [X^\nu,\chi^\lambda,-] +\gamma_i\,
[Y^i,\chi_\mu,-] \big) \ .
\label{twistedsusy}\end{align}
Setting the fermions equal to zero in (\ref{twistedsusy}), one finds
that the corresponding BPS equations for the supersymmetric solutions
of the field theory are
\begin{align}
\nabla_\mu^+ X^\mu-\tfrac{\di}{3}\, \epsilon_{\mu\nu\lambda}\,
[X^\mu,X^\nu,X^\lambda]&=0= \nabla_\mu^+X_\nu-\nabla_\nu^+X_\mu \ ,
\nonumber \\[4pt]
\nabla_\mu^+ Y^i&=0= F_{\mu\nu}^+ \ , \nonumber\\[4pt]
[Y^i,Y^j,Y^k]&=0 = [Y^i,Y^j,X^\mu]
\end{align}
where the twisted field strength is defined by
\begin{align}
F^+_{\mu\nu}: =F_{\mu\nu}-\di\, \epsilon_{\nu\lambda\rho}\,
[\nabla_\mu X^\lambda, X^\rho,-]+\di\, \epsilon_{\mu\lambda\rho}\,
[\nabla^+_\nu X^\lambda, X^\rho,-]
\end{align}
with $F_{\mu\nu}=[\nabla_\mu,\nabla_\nu]$.

\subsection{Mapping to the Blau-Thompson model}

Let us consider the metric 3-Lie algebra $\CA=A_4$ and apply the
Higgsing procedure to the twisted BLG theory.
We proceed by letting the scalar fields $Y^i$ have classical values proportional to
fixed 3-Lie algebra elements. Using $\sSO(5)$ symmetry we can assume
that only $Y^1$ acquires a vacuum expectation value, and by ${\sf SO}(4)$ invariance we can align
this value in the 3-Lie algebra direction $\tau_4$. Hence we make the replacement
\begin{align}
Y^1 \ \longrightarrow \ -g\, \tau_4+Y^1 ~,
\end{align}
where $g$ is a gauge coupling constant. The reduction of the gauge
fields works in the usual way: With respect to the splitting
$\frg_\CA=\aso(4)=\aso(3)\oplus\aso(3)$, we make the replacements
\begin{align}
A^\pm_\mu \ \longrightarrow \ A^\pm_\mu\pm\tfrac{1}{2}\, B_\mu~,
\end{align}
where now we regard $A_\mu^\pm,B_\mu\in \aso(3)$. In the strong
coupling limit $g\to\infty$, 3-brackets containing $Y^1$ reduce to the
brackets
$[X^\mu,X^\nu]:= [X^\mu,X^\nu,-\tau_4]$
of the Lie algebra $\CA'=\aso(3)$; we denote the invariant form on
either factors of $\aso(3)=\asu(2)$ by $\Tr_{\CA'}$, which coincides
with the Cartan-Killing form.
We also define a modified field strength
\begin{align}
\widetilde{F}_{\mu\nu}: =F_{\mu\nu}-\di\,
\epsilon_{\nu\lambda\rho}\, [\nabla_\mu X^\lambda,X^\rho] +\di\,
\epsilon_{\mu\lambda\rho}\, [\nabla_\nu X^\lambda,X^\rho] ~.
\end{align}

By inserting this combination of gauge fields into the total
action $S_{\rm top}+S_{\rm m}$, we find that in the strong coupling limit the field $B_\mu$ only interacts with
the Chern-Simons and scalar kinetic terms algebraically. Its equation of motion reads as
\begin{align}
B_\mu=\tfrac{1}{2g}\, \epsilon_{\mu\nu\lambda}\,
\widetilde{F}{}^{\nu\lambda}-\tfrac{1}{2}\, \nabla_\mu Y^1~,
\end{align}
where we keep only those terms that will remain in the strong coupling
limit. Integrating out the field $B_\mu$, we find that the
Chern-Simons term from the first line of (\ref{BLGtop}) reduces to the
modified Yang-Mills term $\int\, \dd^3x\ \Tr_{\CA'}\big((\widetilde{F}_{\mu\nu})^2 \big)$. One
further finds that the field $Y^1$ decouples from the remaining terms
of the total action, so we introduce a new index
$a=1,2,3,4$. Altogether, after suitable rescaling of the fields we thus find that the reduced action is
given by
\begin{align} \nonumber
S_{\rm red}=&\ \int\, \dd^3x\ 
\Tr_{\CA'}\big(\, \tfrac{1}{2}\, (\widetilde{F}_{\mu\nu})^2
-\tfrac{\di}{2}\, \epsilon^{\mu\nu\lambda}\,
\bar{\chi}_\mu\,(\nabla_\nu\chi_\lambda+[\chi_\nu,X_\lambda]) \\ &
\qquad \qquad \qquad +
\tfrac{1}{2}\, [X^\mu,X^\nu]^2 + [X^\mu,Y^a]^2 +
\tfrac{1}{2}\, [Y^a,Y^b]^2 \nonumber
\\ & \qquad \qquad \qquad +\tfrac{1}{2}\, (\nabla_\mu X^\nu)^2
-\nabla_\mu X^\nu \, \nabla_\nu X^\mu +\tfrac{1}{2}\, (\nabla_\mu
X^\mu)^2 +\tfrac{1}{2}\,
\nabla^\mu Y^a\, \nabla_\mu Y^a \nonumber \\ & \qquad \qquad 
\qquad +\bar{\psi}\,\nabla_\mu
\chi^\mu -\di\,
\bar{\psi}\, [X^\mu,\chi_\mu]-\tfrac{\di}{2}\,
\bar{\psi}\,\gamma_a\, [Y^a,\psi]-\tfrac{\di}{2}\,
\bar{\chi}^\mu\,\gamma_a\, [Y^a,\chi_\mu] \, \big)~.
\label{blauthompson}\end{align}
As with the original twisted BLG action, the reduced action is
the sum of a topological term and a metric dependent cohomological action.

In~\cite{Koh:2009um} it was shown that the Mukhi-Papageorgakis map is compatible
with the topological twisting procedure. We thus expect that the
reduced model is some topological twist of $\mathcal{N}=8$
supersymmetric Yang-Mills theory in three dimensions.  The possible
twists for this gauge theory were classified in \cite{Blau:1996bx}:
One can either arrive at a twisted $\mathcal{N}=2$ supersymmetric
BF-theory, or a twisted $\mathcal{N}=4$ equivariant extension of the
Blau-Thompson model. Comparing our lagrangian \eqref{blauthompson}
with those listed in \cite{Blau:1996bx}, we find that we have obtained
the on-shell formulation of the $\mathcal{N}=4$ equivariant extension
of the Blau-Thompson model; it can be realised as the worldvolume
gauge theory of D2-branes wrapping supersymmetric three-cycles in
Type~IIA string theory. Maximally supersymmetric Yang-Mills gauge
theories on $S^3$ are also considered in~\cite{Fujitsuka:2012wg}.

\subsection{Cohomological IKKT matrix model}

As the equivariantly extended Blau-Thompson model is a twist of
$\mathcal{N}=8$ supersymmetric Yang-Mills theory, its dimensional
reduction should yield some topological twist of the IKKT matrix model. 
The zero-dimensional reduction of the action
(\ref{blauthompson}) becomes
\begin{align}\nonumber
\mathcal{S}_{\rm BT}=& \ \Tr_{\CA'}\big(\, \tfrac{1}{2}\, (
[A_\mu,A_\nu]-\di\,\epsilon_{\nu\lambda\rho}\,
[[A_\mu,X^\lambda],X^\rho]+\di\, \epsilon_{\mu\lambda\rho}\,
[[A_\nu,X^\lambda], X^\rho] )^2 \\ & \qquad \qquad
+\tfrac{1}{2}[A_\mu,X^\nu]^2-[A_\mu,X^\nu]\, [A_\nu,X^\mu]
+\tfrac{1}{2}\, [A_\mu,X^\mu]^2 \nonumber \\ & \qquad \qquad
+ \tfrac{1}{2}\, [X^\mu,X^\nu]^2 +[X^\mu,Y^a]^2+ \tfrac{1}{2}\,
[Y^a,Y^b]^2 +\tfrac{1}{2}\, [A_\mu,Y^a]^2 \nonumber \\ & \qquad \qquad
-\tfrac{\di}{2}\, \epsilon^{\mu\nu\lambda}\, \bar\chi_\mu\,(
[A_\nu,\chi_\lambda]+ [\chi_\nu,\chi_\lambda]) -\tfrac{\di}{2}\,
\bar\chi^\mu\, \gamma_a\, [Y^a,\chi_\mu] \nonumber \\ & \qquad \qquad +\bar\psi\,
([A_\mu,\chi^\mu]-\di\,\bar\psi\, [X^\mu,\chi_\mu]-\tfrac{\di}{2}\,
\bar\psi\, \gamma_a\, [Y^a,\psi])\big) \ .
\label{cohIKKT}\end{align}
This matrix model defines an $\CN=4$ equivariant extension of the
usual IKKT matrix model in ten dimensions, which can be solved
exactly by using localization techniques. It possesses a nilpotent
$\CN=2$ topological symmetry which acts on the fields as
\begin{align}
\delta A_\mu=& \ \bar\eps\, \chi_\mu \ , \nonumber \\[4pt]
\delta X_\mu =& \ \di\, \bar\eps\, \chi_\mu \ , \nonumber \\[4pt]
\delta \chi^\mu =& \ \di\, \epsilon^{\mu\nu\lambda}\, [A_\mu,A_\nu]\,
\eps \ , \nonumber \\[4pt]
\delta\bar\chi_\mu =& \ - \gamma_a\, [A_\mu,Y^a]\, \eps \ , \nonumber
\\[4pt]
\delta\psi =& \ 0 \ , \nonumber \\[4pt]
\delta\bar\psi =& \ -[A_\mu,X^\mu]\, \eps -\di\, \gamma_{ab}\,
[Y^a,Y^b]\, \eps \ , \nonumber \\[4pt]
\delta Y^a =& \ -2\, \di\, \bar\eps\, \gamma^a\, \psi \ .
\end{align}

In \S\ref{3algebramodels} we showed how the reduced ABJM and BLG
models are related. One could thereby hope to lift the cohomological
deformation (\ref{cohIKKT}) of the IKKT matrix model to obtain an
analogous twist of the ABJM matrix model which would enable the exact
computation of the deformed partition function using localization
techniques. However, it was shown in \cite{Koh:2009um} that for three-dimensional
$\mathcal{N}=6$ Chern-Simons-matter theories the only possible twists
involve vector supercharges, and hence it is not possible to directly
obtain such a cohomological deformation of the ABJM theory. In
\S\ref{equiv3algebra} we shall alleviate this problem by constructing
a cohomological matrix model by hand which explicitly localises onto the
BPS equations of the ABJM matrix model; hence it computes an
equivariant index for the model explicitly, and 
moreover possesses the same qualitative features as the matrix model
(\ref{cohIKKT}) under the Mukhi-Papageorgakis map.

\section{Equivariant 3-algebra models\label{equiv3algebra}}

In this final section we shall relate the computation of partition functions
of our supersymmetric quiver matrix models to those of a particular
cohomological matrix model.
Cohomological matrix models comprise a certain type of topological field
theory which are constructed by specifying a set of fields, a set of
equations, and a set of symmetries; the correlation functions
constructed from this data compute intersection numbers on the moduli
space of solutions to the equations modulo the symmetries~\cite{Birmingham:1991ty}. They have actions of the form
\begin{align}
\CS(\mbf\Phi) =\CQ\, \CV(\mbf\Phi)~,
\label{cohCSPhi}\end{align}
where $\CQ$ is the nilpotent BRST charge of the model acting on a gauge-invariant
functional $\CV(\mbf\Phi)$ of the field content $\mbf\Phi$. Matrix models of this type
have appealing properties. For example, they are often
exactly solvable by using localization methods. A prominent example of
this type of theory is due to Moore, Nekrasov and Shatashvili
\cite{Moore:1998et}: They computed the path integral for the
Yang-Mills matrix model by constructing a related cohomological field
theory, and then solving the cohomological deformation using
localization techniques. This formalism was generalised to
a large class of quiver matrix models in~\cite{Cirafici:2008sn}. Since our dimensionally reduced Chern-Simons-matter theories and the IKKT matrix
model are related via the Mukhi-Papageorgakis map, we could expect that the
deformation approach of~\cite{Moore:1998et} can be lifted to our model; in this
section we will apply this approach to the ABJM matrix
model by constructing a related cohomological matrix model and then
computing the path integral using localization methods. The
deformation of the matrix integral is accomplished using the global $\sSU(4)=\sSpin(6)$
R-symmetry of the model, and it preserves $\CN=2$ supersymmetry. It
involves a choice of a generic element in the Cartan subalgebra of the
R-symmetry group, which enables one to
construct well-defined matrix integrals. See~\cite[\S5.1]{Ydri:2012kt} for a relation
between the cohomological deformation of the IKKT matrix model in four
dimensions and the (non-supersymmetric) Chern-Simons matrix model
obtained from dimensional reduction on $S^3$.

\subsection{Equivariant localization}

We begin by summarising the main features involved in
equivariant localization, in a form that we shall employ it. Localization is a technique used in
supersymmetric quantum field theory by which a path integral over an
infinite-dimensional field domain is reduced to a finite-dimensional
integral; here we apply it to reduce the partition functions for our
quiver matrix models to integrals over the critical point locus of some
matrix functional. For this, we perturb the action $\CS(\mbf\Phi)$ of our
model and consider the deformed partition function
  \begin{align}
  \CCZ_t=\int \, \dd\mbf\Phi \ \de^{-\CS(\mbf\Phi)-t \, \CQ\, \CV(\mbf\Phi)}~,
\label{CCZt}\end{align}
where $\dd\mbf\Phi$ is a suitably normalised, supersymmetry-invariant measure on field
space and $t\in\FR$ parameterizes a continuous family of partition
functions such that $\CCZ:= \CCZ_0$ is the partition function of the original
matrix model. Since the action $\CS(\mbf\Phi)$ is supersymmetric, $\CQ\,\CS(\mbf\Phi)=0$, and the
scalar supercharge $\CQ$ is nilpotent on gauge-invariant operators, we have
\begin{align}
\frac{\partial\CCZ_t}{\partial t} =-\int \, \dd\mbf\Phi \ \CQ\,
\CV(\mbf\Phi)\, \de^{-\CS(\mbf\Phi)-t \,\CQ\, \CV(\mbf\Phi)}=-\int \, \dd\mbf\Phi\
\CQ\big(\CV(\mbf\Phi)\, \de^{- \CS(\mbf\Phi)-t \, \CQ\, \CV(\mbf\Phi)} \big)=0~,
\end{align}
where in the last step we have integrated by parts using the
derivation property of the BRST operator $\CQ$ with $\CQ\CS(\mbf\Phi)
=0$, and used invariance of the measure $\dd\mbf\Phi$ on field space
under the BRST symmetry. 
This means that the original partition function $\CCZ=\CCZ_0$ is
computed by (\ref{CCZt}) at any value of $t$. In the limit $t \rightarrow \infty$,
the partition function often simplifies; in particular, if
$\CQ\,\CV(\mbf\Phi)$ is positive definite then the contributions to the
integral in this limit come from the minima $\mbf\Phi_0$ in field space where
$\CQ\,\CV(\mbf\Phi_0) =0$.
The partition function (\ref{CCZt}) can then be evaluated by applying the method of steepest descent.  
The differences between contributions from $\mbf\Phi_0$ and
a generic point $\mbf\Phi$ in field space are exponentially suppressed as
$t\to\infty$; the dominant contributions to this integral therefore
come from points in a neighbourhood $\CCN(\mbf\Phi_0)$ of $\mbf\Phi_0$.  Assuming that
$\de^{-\CS(\mbf\Phi)}$ varies slowly with respect to $\de^{-t\,\CQ\, \CV(\mbf\Phi)}$, the
partition function reduces to
\begin{align}
\CCZ = \int_{\CQ\,\CV(\mbf\Phi)=0}\, \dd\mbf\Phi_0 \ \de^{- \CS(\mbf\Phi_0)} \
\int_{\CCN(\mbf\Phi_0)}\, \dd\mbf\Phi'\ \de^{- t\,
  \CQ\, \CV(\mbf\Phi^\prime\,)}
\label{CCZlocgen}\end{align}
with $\mbf\Phi^\prime\in\CCN(\mbf\Phi_0)$ denoting fluctuations around the
minima $\mbf\Phi_0$; here we have dropped higher order terms using
nilpotency of the supersymmetry variations. The $t$-dependence of the
fluctuation integral in \eqref{CCZlocgen} cancels by supersymmetry of
the measure $\dd\mbf\Phi'$ when one performs the bosonic and fermionic integrations. Note that for
cohomological matrix models with actions of the form (\ref{cohCSPhi}),
we can apply this argument directly to the integral $\int\,
\dd\mbf\Phi\ \de^{-t\,\CS(\mbf\Phi)}$ itself, so that (\ref{CCZlocgen}) is
given by an integral over minima of the original
action~$\CS(\mbf\Phi)$ with $\CS(\mbf\Phi_0)=0$. 

\subsection{Localization of $\CN=2$ Chern-Simons quiver matrix models\label{locN=2}}

Let us apply this formalism to the matrix models having actions \eqref{reducedgeneralchernsimons} with a positive definite
quadratic form $\Tr_\frg$; to ensure convergence of the matrix
integral, here we set $A_0=\di\, A_3$ with $A_3$ hermitean and
$\gamma^0=\di\, \gamma^3$. For the cohomological
deformation of this action we take
\beq
\CQ\CV = \CQ\,\bar\CQ\Tr_\frg\big( \tfrac{1}{2}\, \bar\lambda\, \lambda-2D\, \sigma \big) \ ,
\eeq
where the supercharge $\CQ$ generates the nilpotent supersymmetry
transformations \eqref{genN=2susy} with $\eta=\eps$ and the spinor
normalization $\bar\eps\, \eps=1$. The deformation term then reads
explicitly as
\beq
\CQ\CV = \Tr_\frg\big(-\tfrac{1}{2}\,
[A_\mu,A_\nu]^2-[A_\mu ,\sigma]^2+D^2+\tfrac{\di}{2}\, \bar\lambda\,
\gamma^\mu\, [A_\mu,\lambda]+\di\,[\bar\lambda,\sigma]\, \lambda \big)
\ .
\eeq
Writing $X^I=(A^\mu,\sigma)$, $\Psi=(\lambda^1,\lambda^2)$,
$\Gamma^I=(\gamma^\mu,\di\, \unit)$, and $F=D$, 
this is just the action of the four-dimensional
Yang-Mills matrix model (\ref{4DIKKT}) (with $g=1$). The localization locus
$\CQ\CV=0$ is given by
\begin{align}
[A_\mu ,A_\nu]=0= [A_\mu ,\sigma] \ , \qquad D=0=\lambda=\bar{\lambda}
\end{align}
for the gauge sector, which coincides with the BPS equations (\ref{N=2BPS}). By noting that the matter part of the action
\eqref{CSquivermatter} is
itself a BRST-exact term
\beq
\CS_{\rm m}=\CQ\,\bar\CQ\Tr_\frg\big(\bar\psi\, \psi-2 Z^\dag\, \sigma\,
Z\big) \ ,
\eeq
we may choose the localization locus
\begin{align}
Z=F=0= \psi
\end{align}
for the matter interactions. Then the action
(\ref{reducedgeneralchernsimons}) vanishes at the critical points. For
gauge group $G=\sU(N)$, the fixed point locus thus
coincides with the moduli variety of quadruples $(A_\mu,\sigma)$ of commuting
matrices; for $G=\sU(N_L)\times \sU(N_R)$, it is a subvariety of the
vacuum moduli space of the ABJM matrix model defined by the BPS
equations (\ref{bpsequations}). While the analogous localization procedure works nicely in
the field theory setting to provide exact results for supersymmetric
Chern-Simons-matter theories on
$S^3$~\cite{Kapustin:2009kz,Marino:2011nm} and their dimensional
reductions to a point~\cite{Asano:2012gt,Honda:2012ni}, in our
dimensionally reduced model the result of the localization integral
\eqref{CCZlocgen} comes out to involve terribly divergent integrals
over the Cartan subalgebra of $\frg$ which are beyond
regularization; the partition function in our case is not well-defined
because the action
lacks supersymmetric mass terms for the scalars. Below we shall cure this problem by constructing a cohomological matrix model whose fixed point locus
provides a rigorous definition of the same moduli variety via
a further equivariant deformation parametrized by the R-symmetry group
of the matrix model. We follow the method of~\cite{Cirafici:2008sn} to
compute a supersymmetric
equivariant index using localization techniques. Although
the localization integral still formally diverges, the presence of
twisted masses enables one to define it via a suitable prescription
that we explain in detail.

\subsection{Cohomological matrix model formalism}

As only theories with $\CN\geqslant 4$ supersymmetry can be twisted to
produce deformed scalar supercharges, we focus our attention henceforth
on the $\CN=6$ ABJM matrix model from \S\ref{ABJMmatrix} for
definiteness; we construct a cohomological matrix model which localizes onto the BPS
equations. In view of our discussion from \S\ref{locN=2}, here we consider
instead the localization locus with $A^{L,R}_\mu =0$ as the gauge
fields do not themselves transform under the R-symmetry; the BPS
equations \eqref{bpsequations} then
reduce to the relations \eqref{VZglobalmin} of the double of the ABJM
quiver (\ref{quiverC3Z3}). Put differently, we localize the partition
function of the matrix model onto the F-term constraints rather than
the D-term constraints.
We localize the matrix integral with respect to the equivariant BRST
operator in the gauge
group $G=\sU(N_L)\times \sU(N_R)$, twisted by the toric action of the
maximal torus $\FT^4$ of the R-symmetry group ${\sf SU}(4)$ of
the matrix model; this deforms the nilpotent BRST charge to a
differential of $\sSU(4)$-equivariant cohomology. We denote the generators of this torus by
$\epsilon_i\in\FR$, $i=1,2,3,4$, and set $t_i=\de^{\, \di\, \epsilon_i}$ with the
$\sSU(4)$-constraint 
\beq
\sum_{i=1}^4\, \epsilon_i=0
\label{SU4constraint}\eeq
on the toric parameters. The full symmetry group of the equivariant
model is thus ${\sf U}(N_L)\times {\sf U}(N_R)\times \FT^4
$. The transformation properties of the fields and equations of motion
under the
toric action of $\FT^4$ are given by
\begin{align}
Z^i \ \longmapsto \ \de^{- \di\, \epsilon_i} \, Z^i~, \qquad \CZ^{jk}_i \
\longmapsto \ \de^{-\di\, (\epsilon_j+\epsilon_k-\epsilon_i) }\, \CZ^{jk}_i~.
\end{align}

In order to construct a supersymmetric matrix model we assign
superpartners to these fields to give multiplets $(Z^i,\psi^i)$ with BRST transformations
\begin{align}
\CQ Z^i=\psi^i~, \qquad \CQ\psi^i=\phi_R \, Z^i-Z^i \,
\phi_L-\epsilon_i\, Z^i \ ,
\label{ZpsiBRST}\end{align}
where the hermitian gauge parameters $\phi_{L,R}\in\sEnd(V_{L,R})$ transform in the adjoint
representation of the factors $\sU(N_{L,R})$ of the gauge
group. (There is no sum over $i$ in the second equation.) 
We now add the Fermi multiplet of auxiliary fields
$(\chi^{jk}_i,H^{jk}_i)$ related to the BPS equations, where the
antighosts are defined as maps $\chi_i^{jk}\in\sHom(V_L,V_R)$ with
transformations that read as
\begin{align}
\CQ H^{jk}_i=\phi_R \, \chi^{jk}_i-\chi^{jk}_i \,
\phi_L-(\epsilon_j+\epsilon_k-\epsilon_i)\, \chi^{jk}_i~, \qquad \CQ\chi^{jk}_i=H^{jk}_i~.
\end{align}
To these fields we include the gauge multiplet
$(\phi_{L,R},\bar{\phi}_{L,R},\eta_{L,R})$ which is necessary to close
the BRST algebra off-shell; these fields have transformations
\begin{align}
\CQ \phi_{L,R}=0~, \qquad \CQ \bar{\phi}_{L,R}=\eta_{L,R}~, \qquad \CQ
\eta_{L,R}=\big[\phi_{L,R},\bar{\phi}_{L,R} \big]~.
\end{align}

In order to obtain a localization onto a well-defined moduli space of
matrices that can
be described as a non-singular quotient of a critical locus by the gauge group $G$, we
incorporate additional fields $\varphi,I_{L,R}$ into the collection of
bosonic fields,
together with their superparters $\zeta,\rho_{L,R}$ into the collection
of fermions. 
The new field $\varphi\in\sHom(V_L,V_R)$ transforms in the bifundamental representation
of the ${\sf U}(N_L)\times {\sf U}(N_R)$ gauge group and in the
determinant representation of the R-symmetry, and hence is invariant
under the toric action of $\FT^4$ by (\ref{SU4constraint}).
The fields $I_{L,R}\in V_{L,R}=\FC^{N_{L,R}}$ are also taken to be
invariant under the action of the torus $\FT^4$ for simplicity, and
they transform as vectors under the actions of the left and right
gauge groups $\sU(N_{L,R})$; in what follows we shall refer to the
fundamental matter fields $I_{L,R}$ as
``framing vectors''.
The equations of motion for these additional fields are given by
\begin{align}
\varphi\, I_L =0=  \varphi^\dag \,I_R
\label{varphiIeq}\end{align}
and they ensure stability of the vacua of our quiver matrix model, as
we discuss in detail later on.
Their BRST transformations are
\begin{align}
\CQ\varphi=\zeta~, \qquad \CQ I_{L,R}=\rho_{L,R}~, \qquad
\CQ\zeta=\phi_R\, \varphi-\varphi\, \phi_L~, \qquad
\CQ\rho_{L,R}=\phi_{L,R}\, I_{L,R}~.
\label{stableBRST}\end{align}
We now add the corresponding antighost and auxiliary fields
$\xi_{L,R}\in V_{L,R}^*$ and $h_{L,R}$ with the BRST transformations
\begin{align}
\CQ \xi_{L,R}=h_{L,R}~, \qquad \CQ h_{L,R}=-\xi_{L,R}\, \phi_{L,R}~.
\end{align}
The BRST symmetry $\CQ$ squares to a gauge transformation twisted by a
$\FT^4$ rotation of the fields.

Following the treatment of \S\ref{locN=2}, we will now write down a
cohomological Yang-Mills type matrix model that has this field content, equations
of motion,
and BRST transformations. It
is given by the $\CN=2$ action
\begin{align} \nonumber
\CS_{\rm coh} =&\ \CQ\Tr_V\big(({\chi}^{jk}_i\, ^\dag\, (\tfrac{g}{2} \,
H^{jk}_i-\di\, [Z^j,Z^k;Z_i] )+\psi^i\, (\phi_L\, Z_i^{\dag }-Z_i^{\dag }\,
\bar{\phi}_R)+\eta_L\, [\phi_L,\bar{\phi}_L]-\eta_R\, [\phi_R,\bar{\phi}_R] \\ \nonumber
& \qquad \qquad +\xi_L^\dag \otimes (g^\prime \,h_L-I_R^\dag\, \varphi
)-\xi_R^\dag \otimes (g^\prime \, h_R-I_L^\dag\, \varphi^\dag
)+\bar{\phi}_L\, \rho_L\otimes I_L^\dag - \bar{\phi}_R\,\rho_R\otimes I_R^\dag \\ \nonumber
& \qquad \qquad + (\bar{\phi}_L\, \varphi^\dag -\varphi^\dag\, \bar{\phi}_R)\,
\zeta+
\tfrac{g_1}{2}\,(Z^i\, \psi^\dag _i+Z^\dag _i\, \psi^i)+\tfrac{g_2}{2}\, (I_L\otimes\rho_L^\dag- I_R\otimes \rho_R^\dag) \\ & \qquad \qquad 
+\tfrac{g_3}{2}\, (\varphi\, \zeta^{\dag }+\varphi^{\dag }\,
\zeta)+\mbox{hermitian conjugates} \big)~,
\label{cftmatrixmodel}\end{align}
where we used the canonical identifications
$\sEnd(V_{L,R})=V_{L,R}\otimes V_{L,R}^*$. The deformation by the last
three BRST-exact terms in (\ref{cftmatrixmodel}) removes flat
directions from the matrix integral for the partition function
(see~\cite{Cirafici:2008sn} for details); the equivariant deformation
further has the effect of generating mass terms for all bosonic
fields, which as we will see yields a well-defined matrix integral. Note that the relevant bosonic part of the action from the first line
of (\ref{cftmatrixmodel}) is $\Tr_V \big(\frac{g}2 \,
H^{jk}_i\,^\dag\,H^{jk}_i - \di\,
H^{jk}_i\,^\dag \,\CZ^{jk}_i \big)$; integrating out
$H^{jk}_i$ gives the bosonic potential energy $\frac1{2g} \,
\Tr_V\big(\CZ^{jk}_i\,^\dag \, \CZ^{jk}_i\big)$ and supersymmetry, and
thus the path
integral of the matrix model localizes onto the configurations where
$\CZ^{jk}_i=0$, as desired. 

Since this matrix model is cohomological, it is
independent of the couplings $g,g',g_1,g_2,g_3$ in the action (\ref{cftmatrixmodel}). We can compute the
partition function by taking various limits of these couplings. The
first step is to use the $\sU(N_L)\times \sU(N_R)$ gauge symmetry to
diagonalize the gauge generators $\phi_{L,R}$; we denote their
eigenvalues by $\phi_L^a$, $a=1,\dots,N_L$, and $\phi_R^b$, $b=1,\dots,
N_R$. This change of variables produces
Vandermonde determinants $\prod_{a<b}\, \big(\phi^b_{L}-\phi_{L}^a\big)^2$
and $\prod_{a<b}\,\big(\phi^b_{R}-\phi_{R}^a \big)^2$ in the path
integral measure. Let us now take the limit $g\rightarrow\infty$. The dominant part of the action is
\begin{align}
\tfrac{g}{2} \,\Tr_V\Big({H}^{jk}_i\,^\dag\, H^{jk}_i+{\chi}^{jk}_i\,^\dag\,
\big(\phi_R\, \chi^{jk}_i-\chi^{jk}_i\,
\phi_L-(\epsilon_j+\epsilon_k-\epsilon_i)\, \chi^{jk}_i \big) \Big) ~.
\end{align}
The auxiliary BRST field integrals should not affect the partition
function, so we fix their integration measures such that
\begin{align}
\int \, \dd H^{jk}_i \ \dd H^{jk}_i\,^\dag \ \exp\Big( \Tr_V \big({H}^{jk}_i\,^\dag\, H^{jk}_i\big)\Big)=1~.
\end{align}
Integrating over the fermions gives a factor of the form $\prod_{a,b}\
\prod_{i}\ \prod_{j<k}\, \big(\phi_L^b-\phi_R^
a+\epsilon_j+\epsilon_k-\epsilon_i\big)$. Now we take the limit $g_1\rightarrow\infty$. The relevant part of the action reads as
\begin{align}
g_1\, \sum_{i=1}^4\, \Tr_V\big(\psi^i\, \psi^\dag _i+Z^i\, (Z^\dag _i\, \phi_R-\phi_L\,
Z^\dag _i-\epsilon_i\, Z^\dag _i)\big)~.
\end{align}
Performing the matter integrations puts a term in the localized matrix
integral of the form $\prod_{a,b}\ \prod_i\,
\big(\phi_L^b-\phi_R^a-\epsilon_i-\di\, 0\big)^{-1}$, where we have
added a small imaginary part to the generic real parameters
$\epsilon_i$ to ensure convergence of the gaussian integrations. 
Next we treat the stabilizing fields $I_{L,R},\varphi$ and their
superpartners. We first take the limit $g^\prime\rightarrow\infty$. The dominant part of the action is
\begin{align}
g^\prime\, \Tr_V \big(h_L^\dag\otimes h_L-h^\dag_R\otimes
h_R-\xi^\dag_L\otimes \xi_L\, \phi_L+ \xi^\dag_R\otimes \xi_R\, \phi_R\big)~.
\end{align}
The fields $h^{L,R}$ can be trivially integrated out, while performing
the left and right fermionic integrations puts terms in the path
integral of the form $\prod_a\, \phi_L^a \ \prod_b\,
\phi_R^b$. Finally, performing the path integral in the large $g_2$
limit gives terms of the form
$\big(\prod_a\, \phi_L^a\big)^{-1}\, \big(\prod_b\,
\phi_R^b\big)^{-1}$, while performing the integrations in the limit
$g_3 \rightarrow\infty$ gives terms of the form $\big(\prod_{a,b}\,
(\phi_L^b-\phi_R^a)\big)^{-1}$.

Combining all of the above evaluations, the final result for the
localization of the cohomological matrix integral can be written in
terms of integrations over the left and right gauge generators in the
Cartan torus of the gauge group as
\begin{align}\label{localizedintegral}
\CCZ^{\rm ABJM}_{N_L,N_R}(\epsilon) =&\ \oint\ \prod^{N_R}_{a=1}\, \frac{\mbox{d}\phi^a_R}{2\pi\,\di} \
\prod^{N_L}_{b=1}\, \frac{\mbox{d}\phi^b_R}{2\pi\, \di} \ \frac{
\prod\limits_{a<b}\, \big(\phi_{L}^b-\phi_{L}^a\big)^2 \ \prod\limits_{a<b}\,
\big(\phi_{R}^b-\phi_{R}^a \big)^2}{
\prod\limits_{a=1}^{N_R}\
  \prod\limits_{b=1}^{N_L}\, \big(\phi_L^b-\phi_R^a-\di\, 0\big)}
\\ & \qquad \qquad \qquad \qquad \qquad \qquad \times \ \prod\limits_{a=1}^{N_R}\
  \prod\limits_{b=1}^{N_L}\ \prod_{i=1}^4\, 
\frac{\prod\limits_{j<k}\,
  \big(\phi_L^b-\phi_R^a+\epsilon_i-\epsilon_j-\epsilon_k\big)}{\phi_L^b-
  \phi_R^a-\epsilon_i-\di\, 0}
~.\nonumber
\end{align}
As a Lebesgue integral, this expression formally diverges. Hence we define it via an analytic
continuation to a suitable contour
integral prescription in the complex plane which picks up the poles of
the integrand; the precise choice of contour keeps track of the
auxiliary multiplet of fields that have been eliminated by taking the
large coupling limits above. It is
straightforward to see that the poles occur precisely on the supersymmetric
solutions of the cohomological matrix model. For this, we consider the
critical points of the action (\ref{cftmatrixmodel}) where the fermions are set equal to zero. They are determined
by the zeroes of the BRST charge. By (\ref{ZpsiBRST}) and
(\ref{stableBRST}) the fixed point equations are then
\begin{align} \label{fixedpteqs}
Z^i_{ab}\, \big(\phi_L^b-\phi^a_R-\epsilon_i\big)=0 = \varphi_{ab}\,
\big(\phi_L^b-\phi^a_R\big)=0~, \qquad I^{a}_{R}\, \phi^a_{R}=0= I^{b}_{L}\, \phi^b_{L}
\end{align}
for each $i=1,2,3,4$, $a=1,\dots,N_R$, and $b=1,\dots,N_L$.

We can evaluate the integral (\ref{localizedintegral}) explicitly in dimensions
$N_L=N_R=1$. As its integrand depends only on the combination
$\phi:=\phi_L-\phi_R$ in this case, it can be evaluated from the
residue theorem by picking up the contributions from the simple poles at $\phi=0$ and
$\phi=\epsilon_i$, $i=1,2,3,4$, to get
\begin{align}
\CCZ^{\rm ABJM}_{1,1}(\epsilon) =\prod^4_{i=1}\, \frac{1}{\epsilon_i}\
\prod_{j<k}\, (\epsilon_i-\epsilon_j-\epsilon_k) + \sum^4_{i=1} \,
\frac1{\epsilon_i} \
\prod_{l\neq i}\, \frac{1}{\epsilon_i-\epsilon_l} \
\prod^4_{i^\prime=1} \ \prod_{j<k}\, (\epsilon_{i^\prime} + \epsilon_i-\epsilon_j-\epsilon_k)~.
\end{align}
On the other hand, for $N_R=0$ one finds that the contour integral
vanishes for $N_L\geqslant 5$; more generally, the integral vanishes
for $|N_L-N_R|$ sufficiently large, in agreement with recent analysis
of the ABJM theory through the partition function of the
$\sU(N_L)\times\sU(N_R)$ lens space matrix model~\cite{Awata:2012jb}.
However, for higher dimensions an explicit evaluation of
(\ref{localizedintegral}) becomes increasingly intractable. 

In the
remainder of this section we shall develop an alternative local model
for the fluctuation integrals in \eqref{CCZlocgen} through a geometric
analysis of the neighbourhoods $\CCN(\mbf\Phi_0)$ around the fixed
point subset of the critical point locus with respect to the action of
the R-symmetry torus $\FT^4$. In particular, we compute an equivariant
index
\beq
\CCI_{N_L,N_R}(t) = \Tr_{\CH_{\rm BPS}}\, (-1)^F\,
\prod_{i=1}^4\, t_i^{R_i}
\label{eqindexTr}\eeq
whose infinitesimal limit $\epsilon_i\to0$
explicitly evaluates the contour integrals (\ref{localizedintegral});
here $\CH_{\rm BPS}$ is the Hilbert space of framed BPS states of the
cohomological field theory and $R_i$ are the generators of the Cartan
subalgebra of the global
symmetry group $\sSU(4)=\sSO(6)$. In writing \eqref{eqindexTr} we have
used the fact that the hamiltonian $H$ vanishes in any cohomological
field theory, and set the fugacity $y=1$ for $\sSO(2)$ rotations as we
take $A_\mu=0$.

Because the R-symmetry descends from
the Lorentz group $\sSO(1,10)$ in 11 dimensions, the operators
$t_i^{R_i}$ in the equivariant index rotate the extra
directions $\FC^4$; this suggests that the index (\ref{eqindexTr}) could be
interpreted as the partition function of M-theory compactified on the total space
of an affine $\FR^{10}$-bundle $M$
over $S^1$, with locally flat metric, viewed as a fibration
\beq
\xymatrix{
\FR^2\times\FC^4 \ \ar[r] & \ M \ar[d] \\
 & S^1
}
\eeq
The affine bundle $M$ is defined as the quotient of $\FR^{11}$ by the
$\RZ$-action given by
\beq
\big(x^0\,,\,\vec x\,,\,z^i\big)\ \longmapsto \ \big(x^0+2\pi\, n\,,\, 
\vec x\,,\, \de^{2\pi\, \di\,n\, \epsilon_i}\, z^i\big)
\eeq
for $(x^0,\vec x,z^i)\in
\FR\times \FR^2\times \FC^4$ and $n\in\RZ$. This background has a realization in
11-dimensional supergravity~\cite{NekrasovJapan}
with the global
metric
\beq
\dd s_{11}^2= \big(\dd x^0\big)^2+\dd\vec x\,^2+ \sum_{i=1}^4\,
\big|\dd z^i
-\di\,\epsilon_i\, z^i\, \dd x^0\big|^2 \ .
\label{twistedmetric}\eeq
Since the $\CN=2$ supersymmetry of the cohomological field theory
is not sufficient to fix the R-charges of
gauge-invariant operators, the partition function depends on the
R-charges of the matter fields. 

\subsection{Vacuum moduli space and fixed point analysis}

The partition function (\ref{localizedintegral}) can be regarded as
computing a regularised volume of the non-compact vacuum moduli
space $\frM_{N_L,N_R}$~\cite{Moore:1997dj}, which we now define explicitly. For this, we
recall the equations of motion (\ref{varphiIeq}) which imply that the
vector $I_L$ sits in the kernel and the vector $I_R$ in
the cokernel of $\varphi$. The presence of the bifundmental field
$\varphi$ also implies that the quotient of the
fixed point locus $\CZ_i^{jk}=0$ by the gauge group $G$ is equivalent to a
quotient by the action of the complexified gauge group $G_\FC$. Then the moduli
space can be represented as a quasi-projective variety
\beq
\frM_{N_L,N_R}=\big\{(\CZ_i^{jk})^{-1}(0)\big\}\, \big/\!\!\big/\,
\sGL(N_L,\FC)\times \sGL(N_R,\FC) \ ,
\label{GITquotient}\eeq
where the GIT quotient on the right is taken by removing the points at which the action
of $G_\FC$ is not free. Such a quotient can be defined by imposing an
additional stability condition on the data $(Z^i,I_{L,R},\varphi)$;  a
suitable notion of stability for our purposes can be given as follows:
We say that a datum $(Z^i,I_{L,R},\varphi)$ is stable if there are no
non-trivial proper subspaces $W_{L,R}\subsetneq V_{L,R}$ which contain
the vectors $I_{L,R}$ and which are invariant under the bilinear
commuting operators
$Z_i^\dag\, Z^j$, $Z^j\, Z_i^\dag$ for all $i,j=1,2,3,4$, respectively. Let us demonstrate
that the gauge group $G_\FC$ acts freely on stable data. Suppose that
$(Z^i,I_{L,R},\varphi)$ is fixed by $(g_L,g_R)\in G_\FC$. Then $g_R\,
Z^i= Z^i\, g_L$, $g_L\, Z_j^\dag= Z_j^\dag\, g_R$,  and $g_{L,R}\,
I_{L,R}=I_{L,R}$, which respectively imply that the subspaces
$W_{L,R}=\ker(\unit-g_{L,R})$ have $Z_i^\dag\, Z^j(W_{L})\subset W_L$, $Z^j\,
Z_i^\dag(W_R)\subset W_R$ and
$I_{L,R}\in W_{L,R}$. It follows by stability that
$g_{L,R}=\unit$, and hence the $G_\FC$-action is free. The
corresponding quotient (\ref{GITquotient}) defines a suitable moduli
space of solutions to the BPS equations (\ref{VZglobalmin}) modulo
gauge equivalence.

Let us now characterise the fixed points of this moduli space. A fixed
point $\Pi= (Z^i,I_{L,R},\varphi)\in \frM_{N_L,N_R}^{\FT^4}$ with respect to the action
of $\FT^4\subset\sSU(4)$ is characterized by the condition that an
equivariant rotation is equivalent to a gauge transformation of the fields, so that
\begin{align}
g_R\, Z^i\, g_L^{-1}= t_i^{-1} \, Z^i \ , \qquad g_{L,R}\, I_{L,R}= I_{L,R}
\ , \qquad g_R\, \varphi = \varphi\, g_L \ .
\label{eqgaugeequiv}\end{align}
Under the $\FT^4$-action the vector spaces $V_{L,R}$ admit the weight space decompositions
\begin{align}
V_{L,R}=\bigoplus_{\alpha\in\mathbb{Z}^4}\,
V_{L,R}(\alpha) 
\end{align}
with
\begin{align}\label{weightspacedefinition}
V_{L,R}(\alpha)=\big \{v\in V_{L,R} \
\big| \ g_{L,R}^{-1}\, v=t^{\alpha_1}_1\, t^{\alpha_2}_2\,
t^{\alpha_3}_3\, t^{\alpha_4}_4\, v \big\}
\end{align}
for $\alpha=(\alpha_1,\alpha_2,\alpha_3,\alpha_4)\in\RZ^4$.
It is a straightforward consequence of (\ref{eqgaugeequiv}) that the
nonvanishing components of the maps $(Z^i,I_{L,R}, \varphi )$ are given by
\begin{align}
Z^i\, : \, V_L(\alpha) \ \longrightarrow \ V_R(\alpha-e_i)~, \qquad
I_{L,R}\in V_{L,R}(0)~, \qquad \varphi\,:\, V_L(\alpha) \ \longrightarrow \ V_R(\alpha) \ ,
\label{Zinonvan}\end{align}
where $e_i\in\RZ^4$, $i=1,2,3,4$, is the vector with $1$ in its $i$-th
component and $0$ elsewhere.
With the weight space decompositions \eqref{weightspacedefinition} and
(\ref{Zinonvan}), it is also easy to show that the solution of the
fixed point equations \eqref{fixedpteqs} is given by setting the
eigenvalues of the gauge parameter matrices $\phi_{L,R}$ in this basis
equal to
\begin{align}
\phi_{L,R}^{\alpha^{L,R}}=\sum_{i=1}^4\, \epsilon_i\, \alpha^{L,R}_i~,
\label{phicritical}\end{align}
and $Z^i=0=I_{L,R}$ except for the components
$Z^i_{\alpha-e_i, \alpha}$ and $I_{L,R}^{0}$. Moreover, the only
non-trivial components of the BPS equations (\ref{bpsequations}) are
given by
\beqa
&& Z^j_{\alpha+e_i-e_j-e_k,\alpha+e_i-e_k}\,
\big(Z_i^\dag\big)^{\alpha+e_i-e_k,\alpha-e_k}\,
Z^k_{\alpha-e_k,\alpha} \nonumber \\ && \qquad \qquad \qquad \qquad
\qquad \
= \ Z^k_{\alpha+e_i-e_j-e_k,\alpha+e_i-e_j}\,
\big(Z_i^\dag\big)^{\alpha+e_i-e_j,\alpha-e_j}\,
Z^j_{\alpha-e_j,\alpha} \ ,
\eeqa
and for the conjugates of these equations one has
\beqa
&& \big(Z^\dag_j\big)^{\alpha+e_j+e_k-e_i,\alpha+e_k-e_i}\,
Z^i_{\alpha+e_k-e_i,\alpha+e_k}\,
\big(Z_k^\dag\big)^{\alpha+e_k,\alpha} \nonumber \\ && \qquad \qquad \qquad \qquad
\qquad \
= \ \big( Z_k^\dag\big)^{\alpha+e_j+e_k-e_i,\alpha+e_j-e_i}\,
Z^i_{\alpha+e_j-e_i,\alpha+e_j}\,
\big(Z^\dag_j\big)^{\alpha+e_j,\alpha} \ .
\eeqa

We can describe the graded components of the $\FT^4$-module
decomposition (\ref{weightspacedefinition}) explicitly in terms of the
fixed point
maps as follows. Recalling the discussion at the end of \S\ref{herm3Lie}, we
unambiguously define subspaces of $V_{L,R}$ by
\beq
W_{L}=\bigoplus_{n_{ij}\geq0} \ \prod_{i,j=1}^4\, \big(Z_i^\dag\,
Z^j\big)^{n_{ij}}\, I_L \ , \qquad W_R=\bigoplus_{n_{ij}\geq 0} \
\prod_{i,j=1}^4\, \big(Z^j\, Z_i^\dag\big)^{n_{ij}}\, I_R \ .
\label{WLR}\eeq
Clearly $I_{L,R}\in W_{L,R}$, the subspace $W_L$ is $Z_i^\dag\,
Z^j$-invariant, and $W_R$ is $Z^j\, Z_i^\dag$-invariant for all
$i,j$. Whence $W_{L,R}=V_{L,R}$ by stability, and hence
\beq
V_L(\alpha)=\bigoplus_{\sum_j\, (n_{ij}-n_{ji}) = \alpha_i} \ \prod_{i,j=1}^4\, \big(Z_i^\dag\,
Z^j\big)^{n_{ij}}\, I_L \ , \qquad V_R(\alpha) =\bigoplus_{\sum_j\,
  (n_{ij}-n_{ji})= \alpha_i} \
\prod_{i,j=1}^4\, \big(Z^j\, Z_i^\dag\big)^{n_{ij}}\, I_R \ .
\label{VLRalpha}\eeq
Note that the constraints on the sums in \eqref{VLRalpha} imply that
the weights must satisfy
\beq
\sum_{i=1}^4\, \alpha_i = 0 \ .
\label{alphasum0}\eeq
We define finite sets of lattice points $\Pi_{L,R}\subset\RZ^4$ by
\beq
\Pi_{L,R} = \big\{\alpha\in\RZ^4 \ \big| \ V_{L,R}(\alpha)\neq0 \big\}
\ ,
\label{PiLR}\eeq
with $\big|\Pi_{L,R}\big|= N_{L,R}$ nodes; the meaning of the
restrictions $\Pi_{L,R}\subset\RZ^3$ implied by \eqref{alphasum0} will
be elucidated below. The vertices of these
lattices are related by the actions of commuting
matrices through the commutative diagrams
\beq
\xymatrix@C=30mm{
V_L(\alpha) \ \ar[d]_{Z_k^\dag\, Z^l} \ \ar[r]^{Z_i^\dag\,
    Z^j} & \ V_L(\alpha+e_i-e_j) \ar[d]^{Z_k^\dag\, Z^l} \\
  V_L(\alpha+e_k-e_l) \ \ar[r]_{Z_i^\dag\, Z^j} & \
  V_L(\alpha+e_i+e_k-e_j-e_l)
} 
\eeq
and
\beq
\xymatrix@C=30mm{
V_R(\alpha) \ \ar[d]_{Z^l\, Z^\dag_k} \ \ar[r]^{Z^j\, Z_i^\dag} & \
V_R(\alpha+e_i-e_j) \ar[d]^{Z^l\, Z_k^\dag} \\
V_R(\alpha+e_k-e_l) \ \ar[r]_{Z^j\, Z_i^\dag} & \
V_R(\alpha+e_i+e_k-e_j-e_l)
}
\eeq

We can gain a better combinatorial understanding of the sets
\eqref{PiLR} by employing some machinery from the theory of quiver
representations (see e.g.~\cite{Quivers}); in this setting we identify
torus-invariant framed BPS states
$\Pi$ in the cotangent bundle of the moduli space of framed representations of the ABJM quiver
\eqref{quiverC3Z3} with fixed dimension vector $(N_L,N_R)$. In fact,
many interesting features of BPS states in three-dimensional
supersymmetric gauge theories find natural realisations within the
quiver framework. For example, there is a conjectural Seiberg duality for Chern-Simons gauge theories with
$\CN\geqslant 2$ supersymmetry (see e.g.~\cite{Aharony:2008gk}); in
the present context this duality is
realized as a mutation of quivers, which is a tilting procedure that
therefore yields an equivalence of the corresponding derived
categories of quiver representations~\cite{Vitoria:2007ff}. 

A quiver representation is the same thing as a module for the path
algebra $\CCA$ of the ABJM quiver (\ref{quiverC3Z3}) with relations \eqref{VZglobalmin}. The path algebra $\CCA$ is
generated by acting with arrows $Z^i$, $Z_i^\dag$, $i=1,2,3,4$, on the
framing vectors $I_{L,R}$, as in \eqref{WLR}; we refer to such quiver
representations as cyclic modules. In this setting we replace our
definition of stable points $\Pi$ above with the more natural notion
of $\theta$-stability appropriate to moduli spaces of quiver
representations~\cite{King}. By regarding the conjugate fields
$Z_i^\dag$ as independent arrows, our quiver moduli problem is then
formally equivalent to that of the conifold quiver whose path
algebra is a noncommutative crepant resolution of the conifold
singularity in six dimensions~\cite{Szendroi}, except that we use
multiple framings as in~\cite{Chung} in order to preserve the
left/symmetry inherent in the original ABJM matrix model. This provides us with a concrete
geometrical description of the vacuum moduli space; it would be
interesting to investigate what low-energy brane dynamics this
geometry could
correspond to in this equivariant model.

The R-symmetry torus $\FT^4$ acts on the arrows $Z^i$, $i=1,2,3,4$;
hence it acts on the whole path algebra $\CCA$ and leaves the
relations \eqref{VZglobalmin} invariant. The diagonal torus $\FT^2$ of
the gauge group $G$ induces an action of $\FT=\sU(1)$ on the arrows
via overall rescaling; this can be used to set
e.g. $\epsilon_4=0$. Modding out by this gauge group action, the
overall torus action is $\FT_Q\cong\FT^3$. We shall now argue
that the $\FT_Q$-fixed points are isolated and are parametrized by
certain filtrations of the finite pyramid partitions of the conifold
quiver. For this, we note that the $\FT_Q$-fixed points in the moduli
space of framed cyclic modules correspond bijectively to $\FT_Q$-fixed
ideals in the path algebra $\CCA$. There is a one-to-one
correspondence between $\FT_Q$-fixed modules of the path algebra
$\CCA$ with relations and the $\FT_Q$-fixed annihilator $\ttA$ of
the framing vectors $I_{L,R}\in V_{L,R}$ consisting of stabilising
bifundamental fields $\varphi$ which satisfy \eqref{varphiIeq}; the
finite-dimensional annihilator $\ttA$ is a left ideal of the path
algebra and it is generated by linear combinations of elements of
the same weight. We claim that $\ttA$ is generated by monomials of the
path algebra, such that its class $[\ttA]$ is an isolated $\FT_Q$-fixed
point in the moduli space of cyclic representations with dimension
vector $(N_L,N_R)$. For this, note that $\ttA$ is generated by linear
combinations of path monomials of the same weights. Given a torus
weight $t_1^{\alpha_1}\, t_2^{\alpha_2}\, t_3^{\alpha_3}$, we can find
finitely many monomial paths $p_l$ emanating from the nodes
$V_{L,R}$. Elements of $\ttA$ with weight $t_1^{\alpha_1}\,
t_2^{\alpha_2}\, t_3^{\alpha_3}$ are most generally written as finite
sums of paths $\sum_l\, \xi_l\, p_l$ for some $\xi_l\in\FC$; if
$\xi_{l'}\neq0$, then $p_{l'}$ should be included as one of the
monomial generators of the $\FT_Q$-fixed annihilator $\ttA$, since
each $p_{l'}$ is a linear map from the framing vectors $I_{L,R}$ to
different vector spaces. By exhausting all monomial generators in this
way, we conclude that the torus fixed point $\ttA$ is generated by
monomials and hence corresponds to an isolated point in the moduli
space of quiver representations.

The problem of parametrizing finite-dimensional cyclic $\CCA$-modules
(up to isomorphism) is now equivalent to the problem of parametrizing finite-codimensional
ideals of $\CCA$ (up to $\CCA$-module
isomorphism). Following~\cite{Chung}, they are classified in terms of
filtered pyramid partitions of length two empty room
configurations. Recall~\cite{Szendroi} that a pyramid partition
consists of two types of layers of stones, labelled $L$ (coloured
white) and $R$ (coloured black), which denote one-dimensional
subspaces $V_{L,R}(\alpha)$ of given toric weights $\alpha$ from
\eqref{VLRalpha}. For $i\geq0$, there are $(i+1)^2$ $L$-type stones on
layer $2i$, and $(i+1)\, (i+2)$ $R$-type stones on layer $2i+1$. A
finite subset $\Pi$ of this combinatorial arrangement is a pyramid
partition if, for every stone of $\Pi$, the two stones immediately
above it (of different colour) are also in $\Pi$.

In the ABJM limit $N_L=N_R=N$, we can make this description of the
vacuum moduli space somewhat more explicit. Then the stability
condition implies that the moduli space is a resolution of the $N$-th
symmetric product orbifold \eqref{frMNSymN} provided by the Hilbert
scheme $(\FC^4)^{[N]}$ of $N$ points in $\FC^4$, which parametrizes
zero-dimensional subschemes of $\FC^4$ of length $N$. The map
$(Z^i,I)\mapsto \sum_l\, \lambda_l\, \vec z_l$ from (\ref{ZiSNmap})
gives the Hilbert-Chow map
\beq
\big(\FC^4\big)^{[N]} \ \longrightarrow \ \big(\FC^4\big)^N\, \big/\,
\frS_N
\eeq
which is constructed in detail in~\cite{Jardim}. Following the
derivation in~\cite{Cirafici:2008sn}, the $\FT^4$-fixed
points in this case are parametrized by three-dimensional solid
partitions~\cite{Andrews} of the positive integer $N$; they are
specified by height functions $\Pi(\iota)\in\RZ$ on a
cubic lattice with sites $\iota\in\RZ^3$, such
that $\Pi(\iota)\geq0$ are decreasing functions in each of the three
lattice directions satisfying
\beq
\sum_{\iota\in\RZ^3} \, \Pi(\iota)=N \ .
\eeq

\subsection{Equivariant index for the ABJM quiver}

The localization formula allows one to calculate the contribution to
the partition function from each fixed point; as we have discussed,
the sum over fixed points is captured by applying the residue theorem
to write the contour integral \eqref{localizedintegral} as a sum over
simple poles at the critical points (\ref{phicritical}). As the
explicit form of the residue formula is difficult to handle, we generalize the
technique of~\cite{Bruzzo:2002xf} to
extract the eigenvalues of the superdeterminants of the BRST operator
$\CQ$, arising in the
fluctuation integrals \eqref{CCZlocgen}, from the character of the
tangent space to the moduli space at each critical point. Let $Q$ be the fundamental representation of $\FT^4$
with weight $(1,1,1,1)$; the dual module $Q^*$ has weight
$(-1,-1,-1,-1)$. The local geometry of the moduli space of BPS
solutions $\frM_{N_L,N_R}$
near a particular fixed point $\Pi=(Z^i,I_{L,R},\varphi)$ can be
described by the complex of vector spaces
\begin{equation}
\begin{matrix} \sEnd(V_L) \\ \oplus \\ \sEnd(V_R) \end{matrix} \
\xrightarrow{ \ \dd_1^\Pi \ } \ \begin{matrix} \sHom(V_L,V_R) \ \otimes \ Q \\ \oplus \\
  V_L \ \oplus \ V_R \\ \oplus \\ \sHom(V_L,V_R) \end{matrix} \
\xrightarrow{ \ \dd_2^\Pi \ } \ \begin{matrix} \sHom(V_L,V_R) \ \otimes \ \big(
  Q^*\otimes
  \mbox{$\bigwedge^2 Q$} \big) \\ \oplus \\ V_L \
  \oplus \ V_R \end{matrix}
\label{localcomplex}\end{equation}
where the map $\dd_1^\Pi$ is an infinitesimal gauge transformation 
\begin{align}
\dd_1^\Pi \begin{pmatrix} \phi_L\\ \phi_R\end{pmatrix} =\begin{pmatrix} \phi_R\, Z^i-Z^i\, \phi_L \\
  \phi_L\, I_L \\\phi_R\, I_R \\\phi_R\, \varphi-\varphi\,
  \phi_L\end{pmatrix} \ ,
\end{align}
while the map $\dd_2^\Pi$ is the differential of the equations
\eqref{VZglobalmin} and \eqref{varphiIeq} that define the vacuum moduli space so that
\begin{align}
\dd_2^\Pi \begin{pmatrix} Y^i \\v_L \\v_R
    \\Y \end{pmatrix} = \begin{pmatrix} \big[Y^j,Z^k;Z_i\big] + \big[
    Z^j, Y^k;Z_i\big] + \big[
    Z^j,Z^k;Y_i\big] \\ \varphi^\dag\, v_R+Y^\dag\, I_R \\ \varphi\, v_L
    +Y\, I_L \end{pmatrix} ~.
\end{align}
The first cohomology $\ker(\dd_2^\Pi)/{\rm im}(\dd_1^\Pi)$
parametrizes deformations and provides a local model for the tangent
space $T_\Pi \frM_{N_L,N_R}$ at the fixed point $\Pi$. As
supersymmetric ground states are in one-to-one correspondence with
cohomology classes of $\frM_{N_L,N_R}$, 
the total cohomology of this complex is identified with the Hilbert
space $\CH_{\rm BPS}$ of framed BPS states of the cohomological field theory.

The complex (\ref{localcomplex}) has a natural meaning in the local geometry
of the moduli space of representations of the framed ABJM
quiver. Write $V$ for a given representation of the ABJM quiver
(\ref{quiverC3Z3}) with fixed dimension vector $(N_L,N_R)$, and
$\sExt^p(-,-)$ for the extension groups in the abelian category of
modules for the path algebra $\CCA$. Then the first
term of \eqref{localcomplex} is the space $\sExt^0(V,V)=\sHom(V,V)$ of
nodes of the quiver \eqref{quiverC3Z3}, the second term is the space
$\sExt^1(V,V)$ of arrows including the framing, and the third term is
the vector space $\sExt^2(V,V)$ of all relations; as there are no
relations among the F-term relations \eqref{VZglobalmin}, in our case
$\sExt^p(V,V)=0$ for all $p\geq3$ and the deformation complex contains
only three terms. Note that since here the $\FT^4$ action leaves
invariant the F-term relations \eqref{VZglobalmin} but not the
superpotential \eqref{ABJMsuperpot} itself, the deformation complex
\eqref{localcomplex} is neither symmetric nor self-dual; as a
consequence, the local weight of a fixed point $\Pi$ is not simply a
sign $(-1)^{\dim T_\Pi\frM_{N_L,N_R}}$ but is rather a rational
function of the equivariant deformation parameters $\epsilon_i$,
$i=1,2,3,4$. In the following we compute the equivariant Euler
character of the deformation complex \eqref{localcomplex} for the ABJM quiver. Via our deformation of the nilpotent BRST operator,
the equivariant Euler character can still be interpreted as a Witten
index in the topologically twisted supersymmetric quantum mechanics on
the moduli space $\frM_{N_L,N_R}$ of supersymmetric vacua.

The equivariant character of the complex (\ref{localcomplex}) can be calculated from its
cohomology which is given by an alternating sum of the
weights of the various $\FT^4$ representations. In the representation
ring of the torus group $\FT^4$, one has $Q=\sum_i\,t_i^{-1}$ and
$\bigwedge^2Q=\sum_{i<j}\, t_i\, t_j$, and we obtain the virtual sum
\begin{align}\nonumber
{\sf ch}_\Pi^{\FT^4}(t)=&\ V^\ast_L\otimes V_L+V^\ast_R\otimes V_R
-\Big( \big(V^\ast_L\otimes V_R \big)\, \sum_{i=1}^4\, t_i^{-1}
  +V_L+V_R+V^\ast_L\otimes V_R \Big) \\
&+ \big(V^\ast_L\otimes V_R \big)\, \sum_{i=1}^4\, t_i \ \sum_{j<k}\, t^{-1}_j\, t^{-1}_k +V_L+V_R \ ,
\label{chT4tdef}\end{align}
where we use the weight decompositions of the vector spaces
\begin{align}
V_{L,R}=\sum_{\alpha^{L,R} \in\Pi_{L,R} }\ \prod_{i=1}^4\,
t^{\alpha^{L,R}_i}_i = \sum_{\alpha^{L,R} \in\Pi_{L,R} }\ \prod_{i=1}^3\,
t^{\alpha^{L,R}_i+\alpha^{L,R}_1+ \alpha^{L,R}_2+ \alpha^{L,R}_3}_i
\end{align}
as $\FT^4$ representations, and the second equality here follows from
the constraints \eqref{SU4constraint} and \eqref{alphasum0}; the dual involution acts on the weights as inversion
$(t_i)^*=t_i^{-1}$. Inserting this decomposition into the character
formula (\ref{chT4tdef}) and using the $\sSU(4)$-constraint $t_1\,t_2\,
t_3\, t_4=1$ we find
\begin{align}\nonumber
{\sf ch}_\Pi^{\FT^4}(t)=&\ \Big(\, \sum_{j\neq k}\, t_j\,
t^2_k+2\, \sum^4_{j=1}\, t^{-1}_j -1\, \Big) \ \sum_{\alpha^{L,R}\in\Pi_{L,R}} \
\prod_{i=1}^4\,  t^{\alpha^R_i-\alpha^L_i}_i \\ 
&+ \sum_{\alpha^L,\beta^L\in\Pi_L}\ \prod_{i=1}^4\,
t^{\alpha^L_i-\beta^L_i}_i + \sum_{\alpha^R,\beta^R\in\Pi_R}\
\prod_{i=1}^4\, 
t^{\alpha^R_i-\beta^R_i}_i ~.
\label{chT4expl}\end{align}

The corresponding top form then gives the equivariant version of the fluctuation integral over
the normal bundle $\CCN(\Pi)$ in (\ref{CCZlocgen}) at each fixed point
$\Pi$ of the vacuum moduli space $\frM_{N_L,N_R}$. As the second
cohomology of the complex (\ref{localcomplex}) is non-vanishing, there
is a non-trivial obstruction theory for the moduli space and the
localization formula computes the equivariant Euler character of the virtual tangent
bundle over $\frM_{N_L,N_R}$, i.e. the difference in K-theory between
the tangent and normal bundles at each fixed point of the moduli space. By summing over all fixed points $\Pi$ we arrive at an explicit combinatorial expression for the
contour integral \eqref{localizedintegral} given by the finite sum
\begin{align}
\CCZ^{\rm ABJM}_{N_L,N_R}(\epsilon)=&\ \sum_{\Pi\in\frM_{N_L,N_R}^{\FT^4}}\,
\frac{\prod\limits_{\alpha^L,\beta^L\in\Pi_L}\, \Big(\,
  \sum\limits_{i=1}^4\, \big(\alpha_i^L-\beta_i^L\big)\, \epsilon_i\,
  \Big) \ \prod\limits_{\alpha^R,\beta^R\in\Pi_R}\,\Big(\,
  \sum\limits_{i=1}^4\, \big(\alpha_i^R-\beta_i^R\big)\, \epsilon_i\,
  \Big)}{\prod\limits_{\alpha^{L,R}\in\Pi_{L,R}} \, \Big(\, \sum\limits_{i=1}^4\, \big(\alpha_i^R-\alpha_i^L\big)\, \epsilon_i\,
  \Big)} \nonumber \\ & \times \ \prod\limits_{\alpha^{L,R}\in\Pi_{L,R}} \
  \prod_{j=1}^4\,
  \Big(\big(\alpha_j^R-\alpha_j^L-1\big)\, \epsilon_j+\sum_{i\neq j}\,
  \big(\alpha_i^R-\alpha_i^L\big)\,
  \epsilon_i\Big)^2 \label{CCZvolume} \\ &
  \qquad \qquad \times \ 
\prod_{j\neq k}\, \Big(
  \big(\alpha_j^R-\alpha_j^L+1\big)\,
  \epsilon_j+\big(\alpha_k^R-\alpha_k^L+2 \big)\, \epsilon_k+\sum_{i\neq
    j,k}\, \big(\alpha_i^R-\alpha_i^L\big)\, \epsilon_i \Big) \ . \nonumber 
\end{align}
Consistently with the fact that it computes an equivariant index,
the partition function $\CCZ^{\rm ABJM}_{N_L,N_R}(\epsilon)$ is a Laurent series
in the deformation parameters $(\epsilon_1,\epsilon_2,\epsilon_3)$ with rational coefficients. The
partition weights $\alpha^{L,R}\in\Pi_{L,R}$ in this formula are
naturally interpreted as R-charges of framed BPS particles of the
three-dimensional supersymmetric gauge theory. 

It would be interesting
to match these regularised volumes with those that would appear in
corresponding gravitational free energies for M-theory backgrounds
$AdS_4\times Y_7$ with metrics suitably twisted by the $\sSU(4)$
rotations, as in (\ref{twistedmetric}); see
e.g.~\cite{Herzog:2010hf,Martelli:2011qj,Jafferis:2011zi} for some
examples of field theory computations of (non-equivariant) moduli space volumes in this
context as functions of R-charges. It would also be interesting to see
if there are values of the equivariant deformation parameters for
which our equivariant index compares with the superconformal indices
computed for the full ABJM
theory~\cite{Bhattacharya:2008bja,Kim:2009wb,Yokoyama:2011qu}. Remembering
that the equivariant index of BPS states \eqref{eqindexTr} is properly
defined by compactifying the radial time direction on a circle, it is
computed by the equivariant character of the complex
\eqref{localcomplex} in K-theory (rather than in
cohomology), and from (\ref{chT4expl}) we find explicitly
\begin{align}
\CCI^{\rm ABJM}_{N_L,N_R}(t)=&\ \sum_{\Pi\in\frM_{N_L,N_R}^{\FT^4}}\,
\frac{\prod\limits_{\alpha^L,\beta^L\in\Pi_L}\, \Big(1-
  \prod\limits_{i=1}^4\, t_i^{\beta_i^L-\alpha_i^L} \Big) \
  \prod\limits_{\alpha^R,\beta^R\in\Pi_R}\, \Big(1-\prod\limits_{i=1}^4\,
  t_i^{\beta_i^R-\alpha_i^R}
  \Big)}{\prod\limits_{\alpha^{L,R}\in\Pi_{L,R}} \,
  \Big(1-\prod\limits_{i=1}^4\, t_i^{\alpha_i^L-\alpha_i^R} \Big)}
\nonumber \\ & \qquad \qquad \qquad \times \
\prod\limits_{\alpha^{L,R}\in\Pi_{L,R}} \ \prod_{j=1}^4\,
  \Big(1-t_j^{\alpha_j^L-\alpha_j^R-1} \, \prod_{i\neq
    j}\, t_i^{\alpha_i^L-\alpha_i^R} \Big)^2 \label{CCIindex} \\ &
  \qquad \qquad \qquad \qquad \qquad \times \ \prod_{j\neq k}\, 
\Big(1- t_j^{\alpha_j^L-\alpha_j^R+1} \,
t_k^{\alpha_k^L-\alpha_k^R+2} \, \prod_{i\neq
    j,k}\, t_i^{\alpha_i^L-\alpha_i^R} \Big) \ . \nonumber 
\end{align}
The infinitesimal limit $\epsilon_i\to0$ of (\ref{CCIindex})
reproduces (\ref{CCZvolume}). However, matching with superconformal
index calculations would still require some non-trivial combinatorial identities converting the
sums over partitions into the infinite products (or more precisely
plethystic exponentials) that appear in these formulas.

\subsection{Equivariant index for the Gaiotto-Witten quiver}

A completely analogous calculation of the equivariant index can be
performed for the $\CN=4$ Gaitto-Witten matrix model from
\S\ref{GWmodel} In order to demonstrate the generality of our
formalism, and to justify the constructions of this section, let us briefly explain it. It can be obtained by formally
reducing two of the flavour directions of the
cohomological matrix model for the ABJM model above, which corresponds
to restricting the indices $i,j,k=1,2$ which label the bifundamental
matter fields $Z^i$. In this case we
localize with respect to a two-torus $\FT^2$ which is the maximal
torus of the R-symmetry group $\sSU(2)\times\sSU(2)$; the
corresponding deformation parameters $(\epsilon_1,\epsilon_2)$ are now
independent. For the $\sU(N)\times \sU(N)$ model, the moduli space is
the Hilbert scheme $(\FC^2)^{[N]} \to\frM^{\rm GW}_N$ whose
$\FT^2$-fixed points are parametrized by linear partitions $\lambda$
of the rank $N$. In the case $N_L\neq N_R$, the analogous restrictions
on the vector space sums (\ref{VLRalpha}) imply that there is only one
independent weight $\alpha=\alpha_1=-\alpha_2$ in each of the left and
right sectors of the character lattice $\RZ^2$ of $\FT^2$; whence there is a unique fixed point in each sector
labelled by $\alpha_{L,R}\in\{1,\dots,N_{L,R}\}$ which arises as the
reduction of a filtered pyramid partition along two of its
directions. As a consequence, the partition function will depend only
on the combination $\epsilon_1-\epsilon_2$.

The equivariant deformation complex for Gaitto-Witten quiver is
formally the same as (\ref{localcomplex}), where now
$Q=t_1^{-1}+t_2^{-1}$ and $\bigwedge^2Q=(t_1\,t_2)^{-1}$ in the
representation ring of $\FT^2$, together with the $\FT^2$
characters
\beq
V_{L,R}= \sum_{\alpha_{L,R}=1}^{N_{L,R}} \, \big(t_1\,
t_2^{-1}\big)^{\alpha_{L,R}} = -\, \frac{t_1\,
  t_2^{N_{L,R}}-t_1^{N_{L,R}+1}}{t_1\, t_2^{N_{L,R}}- t_2^{N_{L,R}+1}} \
.
\eeq
The equivariant character is now simply
\beq
{\sf ch}_\Pi^{\FT^2}(t)= V^\ast_L\otimes V_L+V^\ast_R\otimes V_R
-V^\ast_L\otimes V_R
\eeq
and whence the equivariant index of BPS states \eqref{eqindexTr} in
this theory evaluates to
\beq
\CCI^{\rm GW}_{N_L,N_R}(t)=
\frac{\prod\limits_{\alpha_L,\beta_L=1}^{N_{L}} \, \big(1-
  t^{\beta_L-\alpha_L} \big) \
  \prod\limits_{\alpha_R,\beta_R=1}^{N_R} \, \big(1-
  t^{\beta_R-\alpha_R}
  \big)}{\prod\limits_{\alpha_{L}=1}^{N_L} \ \prod\limits_{\alpha_{R}=1}^{N_R} \,
  \big(1-t^{\alpha_L-\alpha_R} \big)} \ ,
\label{CCIindexGW}\eeq
where we defined $t:=t_1\,t_2^{-1}=\de^{-\di\, (\epsilon_1-\epsilon_2)}$.
Note that in the diagonal limit $t_1=t_2$ the index vanishes, thus
demonstrating that our equivariant index indeed gives a more refined
enumeration of degeneracies of BPS states.

The simplicity of the index here as compared to \eqref{CCIindex}
follows from the topological twisting of this $\CN=4$ theory which is
given in~\cite{Koh:2009um}; similarly to \S\ref{BLGtwist}, the
twist is constructed by replacing the Lorentz group $\sSO(3)=\sSU(2)$
with the diagonal subgroup of its product with one of the $\sSU(2)$
factors of the $\sSO(4)=\sSU(2)\times\sSU(2)$ R-symmetry group. With
the twisted BRST charge $\CQ$, the resulting Gaitto-Witten matrix
model action in the absence of gauge fields is proportional to the
$\CQ$-exact term $\CQ\Tr_V\big(\lambda\, \CZ_i^{ij}\big)$ where $\lambda$
is the $\sSO(3)$ scalar part of the twisted spinors $\psi^i$. By deforming this
supercharge suitably as in \eqref{ZpsiBRST}, and adding the
appropropriate quartet of auxilliary fields as before, the
corresponding cohomological matrix integral localises onto a contour
integral over the Cartan subalgebra of the gauge group $\sU(N_L)\times
\sU(N_R)$ which evaluates to \eqref{CCIindexGW}.

\acknowledgements

We thank Michele Cirafici and Christian S\"amann for helpful discussions. This work was partially supported by the Consolidated Grant
ST/J000310/1 from the UK Science and Technology Facilities
Council. The work of RJS was supported in part by Grant RPG-404 from the Leverhulme Trust. 

\bibliographystyle{latexeu}

\end{document}